\documentclass[epj,nopacs]{svjour}

\usepackage{amsmath,amssymb,cite}

\title{Quantum gauge models without (classical) Higgs mechanism}

\author{%
Michael~D\"utsch\inst{1}%
\and
Jos\'e M. Gracia-Bond\'ia\inst{2}%
\and
Florian Scheck\inst{3}%
\and
Joseph C. V\'arilly\inst{4}}

\institute{%
Courant Research Center ``Higher order structures in Mathematics'',
Mathematisches Institut, Univ. G\"ottingen, 37073 G\"ottingen, 
Germany; \email{michael.duetsch@theorie.physik.uni-goettingen.de}
\and
Departamento de F\'isica Te\'orica, Universidad de Zaragoza,
Zaragoza 50009, Spain; \email{jmgb@unizar.es}
\and
Institut f\"ur Physik, Theoretische Elementarteilchenphysik,
Johannes Gutenberg-Universit\"at, 55099 Mainz, Germany;
\email{scheck@uni-mainz.de}
\and
Escuela de Matem\'atica, Universidad de Costa Rica,
San Jos\'e 2060, Costa Rica; \email{joseph.varilly@ucr.ac.cr}}

\date{Received: 18 June 2010 / Revised version: 19 August 2010}

\newcommand{\al}{\alpha}            %% short for \alpha
\newcommand{\bt}{\beta}             %% short for \beta
\newcommand{\dl}{\delta}            %% short for (Dirac's) delta
\newcommand{\eps}{\varepsilon}      %% short for \epsilon
\newcommand{\ga}{\gamma}            %% short for \gamma
\newcommand{\ka}{\kappa}            %% short for \kappa
\newcommand{\La}{\Lambda}           %% short for \Lambda
\newcommand{\la}{\lambda}           %% short for \lambda
\newcommand{\vf}{\varphi}           %% short for \varphi (scalar field)

\newcommand{\Oh}{\mathcal{O}}       %% an ordered product
\newcommand{\Ss}{\mathcal{S}}       %% Schwartz space

\newcommand{\R}{\mathbb{R}}         %% real numbers
\newcommand{\Sf}{\mathbb{S}}        %% S in `S-matrix'

\newcommand{\uno}{\mathbf{1}}        %% unit matrix

\newcommand{\class}{\mathrm{class}} %% classical
\newcommand{\gf}{\mathrm{gf}}       %% gauge-fixing terms
\newcommand{\gh}{\mathrm{gh}}       %% ghostly terms
\newcommand{\kin}{\mathrm{kin}}     %% kinetic term
\newcommand{\T}{\mathrm{T}}         %% time ordering
\newcommand{\total}{\mathrm{tot}}   %% total Lagrangian
\newcommand{\tree}{\mathrm{tree}}   %% tree-level term
\newcommand{\YM}{\mathrm{YM}}       %% Yang-Mills terms

\DeclareMathOperator{\diag}{diag}   %% diagonal matrix
\DeclareMathOperator{\tr}{tr}       %% (matrix) trace

\newcommand{\del}{\partial}         %% short for \partial
\newcommand{\nn}{\nonumber}         %% suppress equation numbering
\newcommand{\otto}{\leftrightarrow} %% 1-1 correspondence
\newcommand{\x}{\times}             %% cartesian product or cross
\newcommand{\7}{\dagger}            %% hermitian conjugate
\renewcommand{\.}{\cdot}            %% scalar product

\newcommand{\Aslash}{{A\mkern-10mu/\,}} %% Yang-Mills field for fermions

\newcommand{\half}{{\mathchoice{\thalf}{\thalf}{\shalf}{\shalf}}}
\newcommand{\quarter}{\tfrac{1}{4}} %% small* fraction  1/4
\newcommand{\shalf}{{\scriptstyle\frac{1}{2}}} %% tiny* fraction  1/2
\newcommand{\thalf}{\tfrac{1}{2}}   %% small* fraction  1/2
\newcommand{\third}{\tfrac{1}{3}} %% small* fraction  1/3
\newcommand{\ut}{{\tilde u}}        %% antighost

\newcommand{\set}[1]{\{\,#1\,\}}    %% set notation
\newcommand{\val}[1]{\langle#1\rangle} %% special value <v>
\newcommand{\word}[1]{\quad\mbox{#1}\quad} %% well-spaced word(s)

\def\wick:#1:{\mathopen:#1\mathclose:} %% Wick-ordered :X X X X:

\begin{document}

\abstract{%
We examine the status of massive gauge theories, such as those usually
obtained by spontaneous symmetry breakdown, from the viewpoint of
causal (Epstein--Glaser) renormalization. The BRST formulation of
gauge invariance in this framework, starting from canonical
quantization of massive (as well as massless) vector bosons as
fundamental entities, and proceeding perturbatively, allows one to
re\-derive the reductive group symmetry of interactions, the need for
scalar fields in gauge theory, and the covariant derivative. Thus the
presence of higgs particles is understood without recourse to a
Higgs(--Englert--Brout--Guralnik--Hagen--Kibble) mechanism. Along the
way, we dispel doubts about the compatibility of causal gauge
invariance with grand unified theories.}

\maketitle

\tableofcontents

\bigskip

\textit{``Sire, je n'avais pas besoin de cette hypoth\`ese-l\`a''}

\rightline{--- Pierre Simon de Laplace}

% \S 1
\section{Introduction}
\label{sec:introibo}

With the start of the LHC operation, the Higgs sector of the Standard
Model (SM) and Higgs' mechanism of spontaneous symmetry breakdown
(SSB) allegedly giving rise to it~\cite{EB,H1,H2,GHK} have become a
topical issue~\cite{PastRemembered}.

The Higgs sector has unsatisfactory aspects, often discussed: the
self-coupling terms appear to be ad-hoc, unrelated to other aspects of
the theory, not seeming to constitute a gauge interaction
\cite[Sect.~22]{AH04}. They raise the hierarchy
problem~\cite{Rafferty,ModernosPragmaticos}. Since most fields
involved in the Higgs mechanism are unobservable, the status question
for it cannot be simply resolved by sighting of higgses%
\footnote{Following L. B. Okun~\cite{Okun}, and for obvious
grammatical reasons, we refer to a (physical) Higgs boson as higgs,
with a lower-case~h.}
in the~LHC. Arguably this is why, at the end of his Nobel lecture,
Veltman wrote: ``While theoretically the use of spontaneous symmetry
breakdown leads to renormalizable Lagrangians, the question of whether
this is really what happens in Nature is entirely
open''~\cite{TiniDixit}. This mistrust, also apparent
in~\cite{TruthInRemembrance}, is more widespread than current
ortho\-doxy would have us believe. The theoretical puzzles, as well as
present phenomenological ones, advise a new look at the scalar sectors
of the~SM and grand unified theories (GUTs) within a quantum field
theoretic framework.

In this introduction we first briefly summarize experi\-mental
information on the Higgs sector of the standard model. Then the
conclusions of the ``reality check'' on the Higgs mechanism worked out
in this paper are compared to the ones of a first reality check
performed in the mid-seventies. Finally we summarize the contents of
the paper.

% \S 1.1
\subsection{Phenomenological puzzles}
\label{ssc:introibo-I}

Generally speaking, precision electroweak measurements were successful
in pinning down new features of the SM and its constituents, even
before actual discoveries took place. Perhaps the best example is
provided by the top quark, whose mass could be estimated rather
precisely from a global analysis of all available electroweak data
before it was found at Fermilab. To illustrate this point we quote a
global fit of SM electroweak data~\cite{GFitter}, excluding the
directly measured top mass, which yields
$178.2 \,\begin{smallmatrix} +9.8 \\ -4.2 \end{smallmatrix}$ GeV. If
the experimental value
$172.4 \,\begin{smallmatrix} +1.2 \\ -1.2 \end{smallmatrix}$ GeV is
included, one obtains the improved fit value
$172.9 \,\begin{smallmatrix} +1.2 \\ -1.2 \end{smallmatrix}$ GeV, not
too different from the indirect determination.

The case of the Higgs boson at present is more complicated. The result
of the standard fit not taking into account direct searches for the
higgs (i.e. the lower limits on the higgs mass~$M_\mathrm{H}$ obtained
at LEP and the Tevatron) is:
\begin{equation}
M_{\mathrm H} = 80 
\begin{smallmatrix} +30 \\[.5\jot] -23 \end{smallmatrix}
\ \mbox{GeV}.
\label{eq:Higgswo} % (1)
\end{equation}
The complete fit of all data, including the lower limits, gives the
estimate
\begin{equation}
M_{\mathrm H} = 116.4 
\begin{smallmatrix} +18.3 \\[.5\jot] -1.3 \end{smallmatrix}
\ \mbox{GeV}.
\label{eq:Higgswi} % (2)
\end{equation}
The deviation of the central values in~\eqref{eq:Higgswo}
and~\eqref{eq:Higgswi} from one another are due to contradictory
tendencies in the data. Most notable among these is the
forward-backward asymmetry $A_\mathrm{FB}^{0,b}$ into $b$-quarks
whose \textit{pull value} in the complete fit is
$$
\frac{1}{\sigma_\mathrm{meas}} \biggl( A_\mathrm{FB}^{0,b}
\Bigr|_\mathrm{fit} - A_\mathrm{FB}^{0,b} \Bigr|_\mathrm{meas} \biggr)
= 2.44,
$$
the denominator $\sigma_\mathrm{meas}$ being the error in the
measurement. This hadronic asymmetry, taken in isolation, yields a
tendency to rather high values of the higgs mass, while the leptonic
asymmetries in the case of the LEP data either agree with the
value~\eqref{eq:Higgswi} or tend towards lower values of that mass as
is the case for the SLD data ---consult Fig.~3 in~\cite{GFitter}. In
spite of the tension between leptonic and hadronic asymmetries, no
single pull value exceeds the $3\sigma$~level. The present state of
such fits, and the influence of the (more than twenty) SM quantities
on them, are well summarized in Table~1 and Figs.~2 and~3
of~\cite{GFitter}. The divergent influences of various electroweak
data on the mass of the higgs were noted already a few years
ago~\cite{Cha-cha-cha}. They have been quantified in
\cite{ELangacker08}, emphasizing the strong correlation between
$A^{(0,b)}_\mathrm{FB}$ and the predicted higgs mass.

The modest quality of the overall fit might be due to inconsistencies
in the data and/or radiative corrections, that might disappear when
further progress is made. However, it might as well signal that the
scalar sector is considerably more complicated than in standard lore,
leading to a reduction of the standard higgs couplings. Consequently
something could have been overlooked at LEP: for instance, mixing with
``hidden world'' scalars~\cite{GJWells08} yields such a reduction, in
particular for the $ZZH$ coupling; and this could not be, and was not,
ruled out by LEP2 for those relatively low energies. Other scenarios
shielding the higgs from detection have been discussed in the
literature ---see \cite{DarkSide,Joraetal} as well as the illuminating
remarks in~\cite{Lightbeing}. Recent experiment has made the situation
even murkier: on Halloween night of 2008, ghostly (albeit rather
abundant) multi-muon events at Fermilab were reported by the CDF
collaboration \cite{AAh,Ptochos}. A possible explanation for them
invokes new light higgs-like particles coupling relatively strongly to
the ``old'' ones, and less so to the SM fermions and vector bosons
\cite{Fermilab,Straggler}. Strassler has cogently emphasized that
``minimality'' of the scalar sector of the SM is just a theoretical
prejudice~\cite{Hamburg}. (Yes, \textit{entia non sunt multiplicanda
praeter necessitatem}. But, who ordered the muon?) The recent
discovery of excess charge asymmetry (that is CP violation) in
$b$-hadrons \cite{NowD0Counterattacks} points in the same
direction~\cite{Dobrescoso}. Given this state of affairs, it seems
premature to draw any definitive conclusion.

% \S 1.2
\subsection{Reality checks for SSB}
\label{ssc:introibo-II}

Causal perturbation theory as developed by H. Epstein and V. Glaser
(EG) and applied to QED by Scharf and collaborators
\cite{DKS,ScharfQED} does not look applicable \textit{a priori} to
non-Abelian gauge theories. Indeed the EG method involves an expansion
in terms of the coupling constant(s) whereas, as is well known, gauge
invariance in the non-Abelian case interrelates terms of different
orders in these couplings. Nevertheless, causal gauge invariance (CGI)
interprets BRST symmetry as a fundamental property of quantum gauge
theory ---in the spirit of~\cite{LuisDixit} and
\cite[Sect.~3.3]{Pokorski}, providing a canonical description of
vector bosons, eliminating unphysical fields, and helping (through the
consistency rela\-tions it imposes) to reconstruct the gauge-invariant
Lagrangian from a general Ansatz.

\smallskip

In this context it may be useful to recall a half-for\-got\-ten
chapter of the early history of gauge theory, chiefly due to Bell,
Cornwall, Levin, Llewelyn Smith, Sucher, Tiktopoulos and Woo in the
seventies: see \cite{CLT74,LT75} and references therein. The
connection between ``tree-unitarity'' (the natural high-energy
boundedness condition for $\Sf$-ma\-trix elements in the tree
approximation) and perturbative unitarity, leading to plausible
renormalizability requisites for Lagrangians, was understood by then.
All those papers started essentially without preconditions from
Lagrangians made out of massive vector bosons (MVB) as fundamental
entities, and found that:

\begin{itemize}
\item
First and foremost, the couplings of the vector bosons had to be of
the gauge theory type, governed by reductive symmetry groups (in
physics parlance, ``groups'' often denote ``Lie algebras'' in this
paper).
\item
Furthermore, scalar fields necessarily entered the picture. The
allowed theories so obtained were essentially equivalent to (the
phenomenological outcome of) SSB models, with one general exception:
``Abelian mass terms'' were possible for the vector bosons. 
\end{itemize}

The latter is understandable: QED with massive photons is a
well-behaved theory. Whenever the symmetry group possesses an
invariant Abelian subgroup, one may add such terms. For our purposes
this second finding is not moot, since it suggests the description of
spin-1 massive models with the help of St\"uckelberg fields. After
all, the SM contains an invariant Abelian subgroup and the mass of the
$Z$ particle can be (though it need not be) thought to be of that
type, see the discussion in Appendix~\ref{app:SM}. St\"uckelberg
fields are of course unphysical. But they have a rightful place in
quantum field theory for the canonical description of~MVB, already at
the level of free fields
\cite{CabezondelaSal,Altabonazo,Felicitas,PCostello}.

\smallskip

Right afterwards the BRST revolution took hold, and the formalism for
gauge symmetry changed forever.%
\footnote{Some of the authors of that reality check openly suspected
SSB as a formal recipe without physical meaning; others were swayed by
the remarkable ``success rate'' of the Higgs mechanism; some
apparently remained agnostic. And so is the case with the present
writers.}

% \S 1.3
\subsection{Outline of the sequel}
\label{ssc:introibo-III}

The book by Scharf~\cite{Scharf}, crowning a successful line of
research \cite{DHKS,DHS,Tobyalone,PGI-EW-I,PGI-EW-II,PepinsFriend}
which in particular establishes a consistent formulation of the SM
without SSB~\cite{PGI-EW-I}, aimed to bring a fresh perspective to the
subject from the standpoint of CGI. In tune with it, with the earlier
reality check, and with the phenomenological SM~Lagrangian, here we
stop pretending we know the origin of mass, and start without
preconditions again from~MVB as fundamental fields. That the reductive
Lie algebra structure then follows from CGI was recognized by Stora
in~\cite{ESItalk}, which constituted an important motivation for this
work.

In Sect.~\ref{sec:CGI-scheme} we expose the theoretical underpinnings
of our own reality check. There are actually at least two CGI methods;
both are expounded there. The first method is constructive. The
second, stemming from a theorem by one of us in~\cite{Michael05}, is
useful rather to \textit{verify} CGI.

Sect.~\ref{sec:mass-patterns} summarizes the first results of the
theory. We report the outcome of that first method, rendering the
cubic coupling relations for CGI models, determined by BRST invariance
at first and second order.

The work on tree-unitarity invoked above seemed to certify
\textit{every} SSB-kind model as acceptable. On the other hand,
Ambauen and Scharf have claimed~\cite{FortunaJuvet} that the CGI
approach clashes with the outcome of SSB for the Georgi--Glashow GUT.
The matter deserved further investigation, all the more so since their
assertion is in contradiction with the second CGI method. As it turns
out, CGI produces constraints on the allowed patterns of masses and
couplings. A certain obstruction put forward by Scharf, sensible
enough in some circumstances, was responsible for the rejection of the
Georgi--Glashow $SU(5)$ and other scenarios. Next in
Sect.~\ref{sec:Sherlock} we unravel this internal problem in CGI by
uncovering an oversight responsible for the mentioned rejections:
there is no problem with GUTs.

Properly reformulated, the obstruction is the germ of the general
$S$-representation, that is, of a derivation from first principles of
the covariant derivative coupling, familiar in the standard
approaches. The theorem in~\cite{Michael05} of course fits with our
construction. The previous analysis allows next to describe what is
presently known to us on quartic terms in the Lagrangian from CGI.

Sect.~\ref{sec:CGI-works}, intended to familiarize the reader
thoroughly with the workings of CGI, is made out of examples. Some
readers might prefer to go to this section before tackling the general
aspects expounded before. First we review an Abelian model. Next we
examine a few slightly more complicated models within CGI. We put
aside the Abelian exception by dealing with simple groups; this
dictates the number of vector bosons (corresponding to Cartan's
classical groups) for irreducible symmetry realizations. We consider
allowed mass patterns for models with one higgs, pondering first the
simple but all-important case with only three gauge bosons, and next
CGI for higher-rank groups. We recall the corresponding SSB mindset:
the choice of only one higgs corresponds to mass patterns produced by
the Higgs mechanism by vector realizations of the gauge group. Then we
look at BRST invariance and minimal coupling from CGI corresponding to
SSB with fields in the adjoint representation.

Sect.~\ref{sec:summa} dwells on our conclusions.

To put matters in perspective, in Appendix~\ref{app:SM} we report
on the SM from the angle of~CGI.%
\footnote{We regret that its elegant formulation without SSB is so
widely ignored.}
Some technical aspects of the machinery underlying this work are
confined to Appendix~\ref{app:constraints}. Finally in
Appendix~\ref{app:GuerraLira} we amplify on the epistemological
implications of the article.

% \S 2
\section{The scheme of causal gauge invariance}
\label{sec:CGI-scheme}

% \S 2.1
\subsection{The method in general}
\label{sec:soyons-cartesiens}

In the origin of causal perturbation theory~\cite{EpsteinGlaser}, the
formulation of gauge symmetry and its preservation in the process of
renormalization was not taken into account. Besides other related
methods to deal with symmetries in that framework
\cite{QNC,MWI,MWIbis,QAP}, causal gauge invariance is a systematic
technique to treat quantum gauge theories perturbatively by
Epstein--Glaser renormalization. As pointed out above, it was first
broached by Scharf and collaborators for QED. A developed formulation
was found in the treatment of massless Yang--Mills
theories~\cite{DHKS}. It has been applied successfully also to massive
non-Abelian models, namely the SM \cite{PGI-EW-I,PGI-EW-II}, spin-2
gauge fields and supersymmetry. Quite recently, the method has been
recruited to examine the issue of the (putative) decoupling of ghosts
in a class of noncovariant gauges~\cite{BurningBurnel}.

Rather than follow the motivation of CGI in the books \cite{ScharfQED}
and~\cite{Scharf}, we adopt a viewpoint inspired by (perturbative)
algebraic quantum field theory. BRST invariance is input already in
the canonical description of vector bosons. The unphysical fields are
eliminated by using the BRST transformation~$s$: the algebra of
observables is obtained as its cohomology, implemented by the
nilpotent BRST charge~$Q$. The space of physical states can be
described cohomologically as well ---see in particular~\cite{DF-QED}.

The construction of $Q$ in perturbative gauge field theory meets the
problem that in general the BRST charge $Q$ changes when the
interaction is switched on~\cite{DF-QED}. For theories with good
infrared behaviour like purely massive theories, Kugo and
Ojima~\cite{KO} showed that $Q$ can be identified with the incoming
(free) BRST charge $Q_\mathrm{in}$, which implements the BRST
transformation $s_0$ of the incoming fields. That the $\Sf$-matrix be
well-defined on the physical Hilbert space of the free theory amounts
to the requirement~\cite{CabezondelaSal,PepinsFriend}:
\begin{equation}
\lim_{g\uparrow 1} \bigl[ Q_\mathrm{in}, \Sf(g\ka L_1) \bigr] 
\Bigr|_{\ker Q_\mathrm{in}} = 0.
\label{eq:good-Smatrix} % (3)
\end{equation}
Here $\Sf(g\ka L_1)$ is the $\Sf$-matrix corresponding to the
interaction $g(x)\ka\,L_1(x)$, an operator in the Fock space of the
incoming free fields. The local Wick polynomial $L_1$ is the part of
the total interaction Lagrangian
$L_\total = \sum_{n=1}^\infty \ka^n L_n$ linear in the coupling
constant~$\ka$ (notationally assumed unique for simplicity). The
function $g \in \Ss(\R^4)$ switches the coupling constant on and off;
the adiabatic limit $g \uparrow 1$ has to be performed to obtain the
physically relevant $\Sf$-matrix. Now, $\Sf(g\ka L_1)$ is a formal
power series,
\begin{align*}
\Sf(g\ka L_1) = \uno + \sum_{n\geq 1} \frac{i^n\ka^n}{n!}
&\int dx_1 \dots dx_n\, g(x_1)\dots g(x_n)
\\
&\quad \x \T_n \bigl( L_1(x_1)\dots L_1(x_n) \bigr),
\end{align*}
where the time ordered product
$\T_n\bigl( L_1(x_1)\dots L_1(x_n) \bigr)$ is an operator-valued
distribution.

The higher order terms of the interaction $L_n$ for $n \geq 2$
---which are also local Wick polynomials--- are taken into account as
local terms in $\T_n\bigl( L_1(x_1)\dots L_1(x_n) \bigr)$: the latter
will contain a term
\begin{equation}
n! (-i)^{n-1}\, \dl(x_1 - x_n, \dots, x_{n-1} - x_n)\, L_n(x_n),
\label{eq:delta-Ln} % (4)
\end{equation}
which propagates to higher orders $n' > n$ by the inductive machinery
of Epstein--Glaser renormalization~\cite{DKS}.

To satisfy \eqref{eq:good-Smatrix} to first order in~$\ka$, one just
searches for a local Wick polynomial $P_\nu$ (called a ``$Q$-vertex'')
such that
\begin{equation}
s_0 L_1(x) \equiv [Q_\mathrm{in}, L_1(x)] = \del^\nu P_\nu(x).
\label{eq:CGI1} % (5)
\end{equation}
Turning to higher orders, we first note that if $x_i \neq x_j$ for all
$i < j$, there is a permutation $\pi$ such that
$x_{\pi(j)} \cap (x_{\pi(j+1)} + V_-) = \emptyset$ for every~$j$, with
$V_-$ being the solid backward lightcone. Hence, for such
configurations the time-ordered product can be written as a standard
operator product:
$$
\T_n\bigl( L_1(x_1) \dots L_1(x_n) \bigr)
= L_1(x_{\pi 1}) \dots L_1(x_{\pi n}).
$$
In view of
\begin{align}
& [Q_\mathrm{in}, L_1(x_1) \dots L_1(x_n)] 
\nn \\
&\quad = \sum_{l=1}^n 
L_1(x_1) \dots [Q_\mathrm{in}, L_1(x_l)]\dots L_1(x_n)
\nn \\
&\quad = \sum_{l=1}^n 
\del_{x_l}^\nu \bigl( L_1(x_1) \dots P_\nu(x_l) \dots L_1(x_n) \bigr),
\label{eq:CGI-motiv} % (6)
\end{align}
one generalizes \eqref{eq:CGI1} to higher orders by requiring that
\begin{align}
& s_0 \T_n(L_1(x_1) \cdots L_1(x_n)) 
\nn \\
&\quad \equiv \bigl[Q_\mathrm{in}, \T_n(L_1(x_1) \dots L_1(x_n))\bigr]
\nn \\
&\quad = \sum_{l=1}^n \del_{x_l}^\nu \,
\T_n\bigl( L_1(x_1) \dots P_\nu(x_l) \dots L_1(x_n) \bigr).
\label{eq:CGIn} % (7)
\end{align}
Formulas \eqref{eq:CGI1} and~\eqref{eq:CGIn} constitute the
\textit{operator} CGI \textit{conditions}, enough to
guarantee~\eqref{eq:good-Smatrix} if the adiabatic limit exists. (In
theories involving massless fields that limit is problematic, to be
sure. For instance, in~QED the $\Sf$-matrix contains infrared
divergences, which cancel in the cross sections. In models with
confinement, the situation is worse, and a perturbative treatment is
possible only for short distances; an adequate description is the
local construction of the observables~\cite{DF-QED} by using couplings
$g(x)\ka$ with a compactly supported test function~$g$. However, the
CGI conditions \eqref{eq:CGI1} and~\eqref{eq:CGIn} are well defined
even in models with bad infrared behaviour; in that case they can be
justified by deriving them from the conservation of the BRST current
for non-constant coupling~\cite{MWI,MWIbis}.)

\smallskip

Requirement~\eqref{eq:CGIn} is a \textit{renormalization condition},
which restricts also tree diagrams, see below. Namely, if the sequence
of time-ordered products $\{T_n\}$ is constructed inductively by
causal perturbation theory~\cite{EpsteinGlaser},
from~\eqref{eq:CGI-motiv} we conclude that CGI can be violated only in
the extension to the total diagonal
$\Delta_n \equiv \set{(x_1,\dots,x_n)\in\R^{4n} : x_1=\cdots=x_n}$ of
the $T_n$; that is, the extension from
$\Ss'(\R^{4n} \setminus \Delta_n)$ to $\Ss'(\R^{4n})$ at the level of
numerical distributions. Indeed there is violation, in that causal
splitting does not respect the divergences in general; however, CGI
can be restored. That gauge-invariant causal renormalization can be
performed to all orders has been proved for QED~\cite{ScharfQED} and
massless $SU(N)$ Yang--Mills theories~\cite{DHKS,DHS}.%
\footnote{One expects that the only obstructions to CGI stem from the
usual anomalies of quantum field theory. Our general ana\-lysis
includes all gauge models which are known to be free of anomalies; but
there are cases where CGI at tree level applied to a general Ansatz
for the Lagrangian generates a model exhibiting anomalies; this
happens for the axial anomaly~\cite{PGI-EW-II}. The question may be
examined quite generally by algebraic criteria developed in
\cite{BHH-I,BHH-II} and~\cite{QAP}. The former authors showed in
particular that gauge-invariant and gauge-fixed cohomologies in the EG
framework are equivalent.}

\smallskip

In summary, within CGI determination by BRST cohomology acts as a
subsidiary physical rule. This answers to the deeply
rooted~\cite{SecundumMathew} need to amend Wigner representation
theory of particles for bosons with spin~1 \textit{in Fock space} with
a Krein structure. Recognition of this is an undoubted merit
of~\cite{Scharf}. (Cherished positivity could be restored at the price
of nonlocality; behind the veil, string-localized potentials and new
field theory phases might well lurk \cite{BS1,BS2}.) We aver that CGI
and the addition of higgs-like fields (in the next subsection) is only
a sufficient and not a necessary condition for unitarity
---consult~\cite{DHS} and~\cite{Oldhand} in this respect. A
nonperturbative understanding of field theory could restore unitarity
in some other way. However, in appropriate contexts and hands,
perturbative methods have something to say about non-perturbative
issues. An example is provided by the exploitation of perturbative
BRST invariance in the understanding of
confinement~\cite{Almasgemelas} ---recently, the same approach has
been applied by Nishijima and Tureanu to the study of the gauge
dependence of the Green's functions in non-Abelian gauge
theory~\cite{AncaN}.

% \S 2.2
\subsection{The grubby machinery}
\label{ssc:gore}

As said above, \eqref{eq:CGI1} and~\eqref{eq:CGIn} are already
nontrivial when used for tree diagrams. CGI strongly restricts the set
of allowed models, determining the interaction
$L = \sum_{n=1}^\infty \ka^n L_n$ to a great extent, independently of
the infrared behaviour. Given the free theory, one makes a polynomial
and renormalizable Ansatz for~$L_1$. The CGI condition~\eqref{eq:CGI1}
determines most of the coefficients in this Ansatz, or yields
relations between them. Turning to higher-order tree diagrams, terms
of the form~\eqref{eq:delta-Ln} remain undetermined in the inductive
Epstein--Glaser construction. Then \eqref{eq:CGIn} determines the
higher-order interaction terms $L_n$ and the as yet undetermined
coefficients of~$L_1$. The process is constructive. The reductive Lie
algebraic structure and the need to add additional physical scalar
fields (higgs fields) in massive non-Abelian models are not to be put
in; they follow from CGI. To be precise in the last respect: for such
a model with MVB and only the unphysical fermionic ghosts and
St\"uckelberg fields, CGI breaks down at order~$\ka^2$
\cite{PGI-EW-I}; however the inclusion of at least one additional
physical scalar makes CGI solvable. The ``puzzling''
\cite[Preface]{DogDine} existence of fundamental scalars is demanded
in our framework.

The process terminates after a finite number of steps in
renormalizable theories. Consider spin-1 gauge models in 4-dimensional
Minkowski space with an $L_1$ trilinear in the fields whose mass
dimension is~$\leq 4$. Then tree diagrams can give nontrivial
constraints only \textit{up to third order}. Indeed, because CGI can
be violated only in the extension of $T_n$ to $\Delta_n$, a possible
violation of~\eqref{eq:CGIn} must be of the form
$$
\sum_{a,\Oh} C_{a,\Oh} \,\del^a\dl(x_1 - x_n, \dots, x_{n-1} - x_n)
\,\Oh(x_1,\dots,x_n),
$$
where $\Oh(x_1,\dots,x_n)$ denotes a normally ordered product of free
fields, $a = (a^\mu_l)^{\mu=0,1,2,3}_{l=1,\dots,n-1}$ is a multi-index
and the $C_{a,\Oh}$ are suitable numbers. Power counting yields the
restriction $|a| + \dim\Oh \leq 5$ with
$|a| \equiv \sum_{l,\mu} a_l^\mu$. Since each vertex of $L_1$ has
three legs, it follows that a tree diagram to $n$-th order satisfies
$\dim\Oh \geq 2 + n$. Combining these two inequalities, one sees that
CGI can be violated at tree level only for $n \leq 3$. In practice,
most interesting information is concentrated at the first and second
orders; third-order CGI only refines a few coefficients in the Higgs
sector.

We illustrate the construction of the time-ordered products at the
tree level at second order, in the case of spin-1 gauge fields,
following essentially \cite{PGI-EW-I}; the bulk of calculations
buttressing this paper are of this type. Note
\begin{align*}
T_2(x,y)\bigr|_\tree
&:= \T_2\bigl(L_1(x)L_1(y)\bigr)\bigr|_\tree,
\\
T_{2/1}^\nu(x,y)\bigr|_\tree
&:= \T_2\bigl( P^\nu(x) L_1(y) \bigr)\bigr|_\tree,
\end{align*}
as well as $T_{2/2}^\nu(x,y)\bigr|_\tree =
T_{2/1}^\nu(y,x)\bigr|_\tree$, can be written as
$$
T_2\bigr|_\tree = T_2\bigr|_\tree^0 + N_2, \qquad
T_{2/1}^\nu\bigr|_\tree = T_{2/1}^\nu\bigr|_\tree^0 + N_{2/1}^\nu,
$$
where $T_2\bigr|_\tree^0$, $T_{2/1}^\nu\bigr|_\tree^0$ include all
terms not vanishing for $x \neq y$; these are the terms with the
Feynman propagator $\Delta_m^F$ or its derivatives
$\del_\mu \Delta_m^F$, $\del_\nu \del_\mu \Delta_m^F$. We replace
$\square \Delta_m^F$ by $-m^2\,\Delta_m^F + \delta$; the
$-m^2\,\Delta_m^F$ term belongs to $T_2\bigr|_\tree^0$ and the
$\delta$-term to~$N_2$. Expressions $N_2$ and $N_{2/1}$ are of the
form
$$
N_2(x,y) = \ka^2\sum_{\phi_1\phi_2\phi_3\phi_4}
C_{\phi_1\phi_2\phi_3\phi_4}\,\dl(x - y)\,\wick:\phi_1\phi_2\phi_3
\phi_4(x):
$$
and similarly for $N_{2/1}$, where $C_{\phi_1\phi_2\phi_3\phi_4}$ are
$c$-numbers, and the sum runs over all kinds of free fields
$\phi_1, \dots,\phi_4$ present in the model. In the framework of
causal perturbation theory, such local terms may be added to~$T_2$, if
they respect power counting, Lorentz covariance, unitarity, ghost
number, etc. A~glance at formula~\eqref{eq:delta-Ln} indicates that
$N_2(x,y) = -2i\,\dl(x - y)\,L_2(x)$, where $L_2$ is a sum of quartic
terms.

In contrast to $T_2\bigr|_\tree^0$ and $T_{2/1}^\nu\bigr|_\tree^0$,
which are uniquely given in terms of $L_1$ and~$P^\nu\!$, the
coefficients in $N_2$ and~$N_{2/1}$ are not yet determined. To prove
CGI for second order tree diagrams, we have to show that the as yet
undetermined coupling parameters of~$L_1$ and the coefficients
$C_{\phi_1\phi_2\phi_3\phi_4}$ in $N_2$ and~$N_{2/1}$ can be chosen in
such a way that
\begin{align}
& \bigl[Q_\mathrm{in}, \bigl(T_2\bigr|_\tree^0 + N_2\bigr)(x,y)\bigr] 
\nn \\
&\quad = \del_\nu^x\bigl(T_{2/1}^\nu\bigr|_\tree^0
+ N_{2/1}^\nu \bigr)(x,y) + [x \otto y].
\label{eq:CGI-2} % (8)
\end{align}
Since this condition holds by induction for $x \neq y$, we need only
study the local contributions. However, the splitting of a
distribution into local and nonlocal parts in principle is not unique,
and some caution is called for. Recall that we replace
$(\del)\square \Delta_m^F$ by $-m^2(\del)\Delta_m^F + (\del)\delta$.
Then for $x \neq y$ only terms $\sim \Delta_m^F$,
$\sim \del_\mu \Delta_m^F$, $\sim \del_\nu \del_\mu \Delta_m^F$ and
$\sim \del_\nu \del_\mu \del_\la \Delta_m^F$ with no contraction of
Lorentz indices contribute to~\eqref{eq:CGI-2}. Since these terms
cancel for $x \neq y$, they cancel for $x = y$ also. There remain only
terms $\sim \dl(x - y)$ and $\sim \del\dl(x - y)$. For spin-1 gauge
theories, such $(\del)\dl$-terms can be generated only in the
following ways.
\begin{enumerate}
\item % 1.
First of all,
\begin{align*}
& [Q_\mathrm{in}, N_2(x,y)] 
= \ka^2 \sum_{\phi_1\phi_2\phi_3\phi_4} C_{\phi_1\phi_2\phi_3\phi_4}
\,\dl(x - y) \ \x
\\
& \bigl(\wick: [Q_\mathrm{in},\phi_1(x)]\,\phi_2(x) \cdots:
+ \wick: \phi_1(x)\,[Q_\mathrm{in},\phi_2x)] \cdots: +\cdots \bigr)
\end{align*}
\item % 2.
{}From $N_{2/1}^\nu(x,y) = C^\nu\,\dl(x - y)\,M(x)$, where
$M = \wick:\phi_1\phi_2\phi_3\phi_4:$, we obtain
\begin{align}
& \del_\nu^x N_{2/1}^\nu(x,y) + [x \otto y]
\nn \\
&\quad = C^\nu \,\del_\nu^x \bigl( \dl(x - y)\,M(x) \bigr)
+ [x \otto y]
\nn \\
&\quad = C^\nu \,\dl(x - y) \,\del_\nu M(x).
\label{eq:did1} % (9)
\end{align}
\end{enumerate}

In addition, $\del_\nu^x T_{2/1}^\nu\bigr|_\tree^0(x,y) + [x \otto y]$
also contains some $(\del)\dl$-terms, generated due to the propagator
equation $(\square + m^2)\Delta^F_m = \dl$:
\begin{enumerate}
\addtocounter{enumi}{2}
\item % 3.
If $P^\nu = b\,\wick:\del^\nu \phi\,F: +\cdots$ and
$L_1 = a\,\wick:\phi\,E: +\cdots$, then the contraction of
$\del^\nu \phi(x)$ with $\phi(y)$ gives a propagator
$-i\del^\nu \Delta^F_m(x - y)$, and on computing its divergence we
find the contribution
\begin{align*}
& \del_\nu^x T_{2/1}^\nu\bigr|_\tree^0(x,y) + [x \otto y] 
\\
&\quad = -2iba\,\dl(x - y)\, \wick:F(x)\,E(x): +\cdots \,.
\end{align*}
\item % 4.
With $P^\nu$ as before and
$L_1(y) = a_\mu\,\wick:\del^\mu\phi(y) \,E(y): +\cdots$ the
contraction of $\del^\nu \phi(x)$ with $\del^\mu\phi(y)$ gives a
propagator $i\del^\nu \del^\mu \Delta^F_m(x - y)$. On computing the
divergence we now obtain a $\del\dl$-term, which we transform into a
$\dl$-term by using the following identity
$$
M(y) \,\del^x_\nu \dl(x - y)
= M(x) \,\del^x_\nu \dl(x - y) + \del_\nu M(x) \,\dl(x - y);
$$
so that
\begin{align}
& \del_\nu^x T_{2/1}^\nu\bigr|_\tree^0(x,y) + [x \otto y] 
\nn \\
&= iba_\mu\,\del^\mu\dl(x - y)\,\wick:F(x)\,E(y): + [x \otto y] +\cdots
\nn \\
&= iba_\mu\,\dl(x - y) \bigl( \wick:F(x)\,\del^\mu E(x):
- \wick:\del^\mu F(x)\,E(x): \bigr) 
\nn \\
&\quad +\cdots \,.
\label{eq:did2} % (10)
\end{align}
\end{enumerate}

After the transformations \eqref{eq:did1} and~\eqref{eq:did2}, all
terms remaining in \eqref{eq:CGI-2} are of the form
$c_1 \,\dl(x-y)\, \wick:\del\phi_1\, \phi_2 \phi_3 \phi_4(x):$ or
$c_2 \,\dl(x-y)\, \wick:\phi_1 \phi_2 \phi_3 \phi_4(x):\,$. These are
linearly independent. Therefore \eqref{eq:CGI-2} is equivalent to a
system of $c$-number equations, obtained by equating the coefficients
belonging to the same Wick monomial.

% \S 2.3
\subsection{The second CGI method}
\label{ssc:CHI-II}

Basic model-building according to CGI uses tree-diagram calculations.
In this connection, and alternatively to the previous method, models
satisfying CGI at tree level can also be obtained by using that
\textit{classical} BRST invariance of the Lagrangian \textit{implies}
CGI for tree diagrams to all orders~\cite{Michael05}. The ideas behind
this can be summarized thus: given a BRST-invariant free theory, that
is
$$
s_0L_0 = \del_\mu I_0^\mu =: \del\.I_0 \word{for some local $I_0$,}
$$
with $L_0$ quadratic in the fields, seek deformations
$L_0 \to L_\total = \sum_{n=0}^\infty \ka^n L_n$ and
$s_0 \to s = \sum_{n=0}^\infty \ka^n s_n$ (with $L_n$, $s_n$
satisfying some obvious properties), such that
\begin{equation}
s L_\total = \del\. I
\word{where}  I^\mu = \sum_{n=0}^\infty \ka^n I^\mu_n
\label{eq:sL} % (11)
\end{equation}
is some local power series. Here $L_\total$ is assumed to contain only
first-order derivatives; have a look at~\cite{LuisDixit}. BRST
invariance of the Lagrangian in this sense implies CGI for tree
diagrams to all orders by the following: in the case of a constant
coupling $\ka$ formula~\eqref{eq:sL} implies conservation of the
corresponding classical Noether (BRST) current:
$\del\. j_{\ka,\class} = 0$. Replacing $\ka$ by $\ka g$, for a test
function~$g$, a generalized current conservation can be derived
from~\eqref{eq:sL}:
\begin{equation}
\del\. j_{\ka g,\class}(x) = \del g(x) \. P_{\ka g,\class}(x),
\label{eq:currcons} % (12)
\end{equation}
where $P_{\ka g,\class}(x)$ is that classical interacting vector field
which agrees for $\ka = 0$ with the $Q$-vertex $P$ of \eqref{eq:CGI1},
more precisely with the corresponding classical (local) field
po\-lynomial. Since the classical limit of an interacting perturbative
quantum field is given by the contribution of connected tree diagrams,
current conservation~\eqref{eq:currcons} can be expressed as a tree
diagram relation in perturbative quantum field theory. Smearing out
this relation with suitable test functions, conservation of the BRST
current goes over to commutation with the free BRST charge
$Q_\mathrm{in}$ belonging to the conserved Noether current of the
symmetry $s_0 L_0 = \del\.I_0$. In this way the CGI
relation~\eqref{eq:CGIn} for tree diagrams to all orders is obtained.

For theories with MVB one can use that, generically, models coming
from SSB are classically BRST invariant in the sense of~\eqref{eq:sL};
this will be spelled out in Sect.~\ref{ssc:suerte-o-verdad}.
Therefore they satisfy CGI at tree level. Most likely, the two methods
outlined here are equivalent, in the sense that the sets of allowed
models are the same. It remains that, whereas the first method amounts
to a direct (perchance tedious) search for the general solution of the
CGI conditions for tree diagrams, we do not know whether the second
yields the most general solution as well.

% \S 3
\section{Mass and interaction patterns}
\label{sec:mass-patterns}

Consider a model with $t$ intermediate vector bosons $A_a$ in all, of
which any may be in principle massive or massless. Let there be $r$
massive ones ($a = 1,\dots,r$, with masses $m_a > 0$) and $s$ massless
($a = r+1,\dots,r+s$), so $t = r + s$. They are accompanied by $z$
physical scalar particles $\vf_p$ of respective masses~$\mu_p$. The
free BRST transformation $s_0 \equiv[Q_\mathrm{in},\cdot]_\mp$ is a
superderivation commuting with partial derivatives, hence given by its
action on the basic fields:
\begin{gather*}
s_0 A^\mu_a = \del^\mu u_a,  \quad
s_0 B_a = m_a u_a,  \quad  s_0 u_a = 0,
\\
s_0 \ut_a = -(\del\. A_a + m_a B_a),  \quad  s_0 \vf_p = 0.
\end{gather*}
Here we let $B_a$ denote the St\"uckelberg field associated to the
vector field $A_a$; in case $A_a$ is massless, $B_a$ drops out. The
total bosonic interaction Lagrangian is of the form
$L_\mathrm{int} = \ka L_1 + \ka^2\,L_2$. For $L_1$ make the following
Ansatz (with unknown coefficients $f^*_{***}$ in the terms below). Let
$L_1 = L_{1B} + L_{1\vf}$, the higgs-free cubic couplings being
$L_{1B} = L_1^1 + L_1^2 + L_1^3 +L_1^4$ and $L_{1\vf} =
L_1^5 + L_1^6 + L_1^7 + L_1^8 + L_1^9 + L_1^{10} + L_1^{11}$ being the
couplings involving physical scalars. In the Feynman gauge, the
allowed higgs-free cubic couplings are
\begin{align}
L_1^1 &= f_{abc} \bigl[ A_a \. (A_b \. \del) A_c
- u_b (A_a \. \del\ut_c) \bigr];
\nn \\
L_1^2 &= f^2_{abc} (A_a\.A_b) B_c;
\nn \\
L_1^3 &= f^3_{abc} \bigl[ (A_a \. \del B_c) B_b
- (A_a \. \del B_b)B_c \bigr];
\nn \\
L_1^4 &= f^4_{abc} \ut_a u_b B_c;
\label{eq:higgfree} % (13)
\end{align}
and the remaining ones, involving higgses, are
\begin{align}
L_1^5 &= f^5_{abp}
\bigl[(A_a \. \del\vf_p) B_b - (A_a \. \del B_b) \vf_p \bigr];
\nn \\
L_1^6 &= f^6_{aqp}
\bigl[ (A_a \. \del\vf_p) \vf_q - (A_a \. \del\vf_q) \vf_p \bigr];
\nn \\
L_1^7 &= f^7_{abp} (A_a \. A_b) \vf_p; 
\nn \\
L_1^8 &= f^8_{abp} \ut_a u_b \vf_p;
\nn \\
L_1^9 &= f^9_{abp} B_a B_b \vf_p;
\nn \\
L_1^{10} &= f^{10}_{apq} B_a \vf_p \vf_q;
\nn \\
L_1^{11} &= f^{11}_{pqr} \vf_p \vf_q \vf_r.
\label{eq:higgful} % (14)
\end{align}
As products of field operators, these monomials are understood to be
normally ordered. Some symmetry relations of the coefficients under
exchange of indices are evident from the definition. Because the
dimension of the Lagrangian must be $M^4$ in natural units, and the
boson field dimension is~1 in our formulation, the coefficients
$f, f^3, f^5, f^6$ are dimensionless, and
$f^2, f^4, f^7,\dots, f^{11}$ have dimension of mass. It is taken into
account that CGI holds a term in $B_a B_b B_c$ to vanish.
With that, the formulas \eqref{eq:higgfree} and~\eqref{eq:higgful} give
the most general trilinear and renormalizable Ansatz modulo divergence
terms and $s_0$-coboundaries. 

We list the determination of the couplings in terms of the $f_{abc}$
and the pattern of masses imposed by CGI at orders $\ka$ and~$\ka^2$,
still essentially in the version of~\cite{Scharf}.

\begin{enumerate}
\item % 1.
As repeatedly indicated, and like in Sect.~\ref{ssc:introibo-II},
with independence of the masses CGI unambiguously leads to gauge
fields with real coupling parameters $f_{abc}$ that are totally
antisymmetric and satisfy the Jacobi identity: that is, to generalized
Yang--Mills theories on reductive Lie algebras. This is remarkable.

\item % 2.
When all $A_a$ are massless, there is no need to add physical or
unphysical scalars for renormalizability, and only $L_1^1$ (and later,
the quartic coupling $L^1_2$) survive. They of course coincide
respectively with the first- and second-order part of the usual
Yang--Mills Lagrangian. In particular: CGI gives rise to gluodynamics.
 
\item % 3.
The relation
\begin{equation}
2 m_c f^2_{abc} = (m_b^2 - m_a^2) f_{abc}
\label{eq:madre-del-cordero} % (15)
\end{equation}
holds. Thus if $m_c = 0$ and $f_{abc} \neq 0$, then $m_a = m_b$
necessarily. And if $m_c \neq 0$, then
$$
f^2_{abc} = f_{abc} \frac{m_b^2 - m_a^2}{2m_c}.
$$
The useful relation between masses and structure constants:
\begin{equation}
(m_b^2 - m_a^2) \sum_{c:\,m_c=0} (f_{abc})^2 = 0;
\label{eq:the-thin-edge} % (16)
\end{equation}
follows directly from~\eqref{eq:madre-del-cordero}. 

\item % 4.
The relation
\begin{equation}
2(f^3_{bca} m_a - f^3_{acb} m_b) = f_{abc} m_c
\label{eq:any-wool} % (17)
\end{equation}
holds. From this, after multiplication by $m_c$ and cyclic
permutation, one obtains the important formula
\begin{equation}
f^3_{abc} = f_{abc}(m_b^2 + m_c^2 - m_a^2)/4m_bm_c.
\label{eq:little-lamb} % (18)
\end{equation}
If either $m_b$ or~$m_c$ vanishes, then $f^3_{abc} = 0$.

\item % 5.
$f^4_{abc} = f_{abc}(m_c^2 - m_b^2 + m_a^2)/2m_c$.

\item % 6.
In the non-Abelian case, when some $A_a$ are massive, coefficients
$f^5$ to $f^{11}$ cannot all vanish: renormaliza\-bility asks for
physical Higgs bosons. The $L_1^5$ and $L_1^6$ terms are the nub of
the problem. Reference~\cite{Scharf} claims that $L_1^6$ is just zero
and that the coefficients of $L_1^5$ are \textit{diagonal} in the
sense that $f^5_{abp} = C_{5p}m_a\,\dl_{ab}$, where the $C_{5p}$ (with
dimension $M^{-1}$) are independent of~$a$; but this is only warranted
when there is a single higgs field, for which the $L_1^6$ term is
absent. A relatively involved expression, given in the next
subsection, ties this key coupling with the structure constants and
the masses.

\item % 7.
$f^7_{abp} = - f^8_{abp} = m_b f^5_{abp}$. This is $C_5
m_a^2\,\dl_{ab}$ when $z = 1$.

\item % 8.
Because $f^7$ is obviously symmetric in the first two indices, so too
is~$f^8$. Now, the symmetry for $f^8$ implies
\begin{equation}
m_b\, f^5_{abp} = m_a\, f^5_{bap}.
\label{eq:missing-link} % (19)
\end{equation}
Note that $f^5_{abp} = 0$ if $m_a = 0$ or $m_b = 0$: for $m_b = 0$
this is clear (no St\"uckelberg field $B_b$), and for $m_a = 0$ it
follows from~\eqref{eq:missing-link}.

\item % 9.
$f^9_{abp} = -(\mu_p^2/2m_a) f^5_{abp}$ for $m_a > 0$;
$f^9_{abp} = 0$ if $m_a = 0$. This is $-\half C_5 m_H^2 \dl_{ab}$ when
$z = 1$, with $\mu_1 \equiv m_H$.

\item % 10.
$f^{10}_{apq} = \frac{\mu_q^2 - \mu_p^2}{m_a} f^6_{apq}$ for $m_a >
0$, with $f^{10}_{apq} = 0$ if $m_a = 0$. This vanishes when $z = 1$.

\item % 11.
The $f^{11}_{pqr}$ are not determined, except (with the help of
third-order tree graphs) in the case of only one higgs; then $f^{11} =
-\half C_5 m_H^2$.
\end{enumerate}

% \S 3.1
\subsection{The first CGI parameter constraint}
\label{ssc:ties-that-bind}

In the preceding subsection we have listed all conditions coming from
CGI for first and second-order tree diagrams which determine directly
the coupling parameters $f^*_{***}$ in terms of $f_{abc}$ and the
masses. However, for second-order tree diagrams CGI gives further
constraints relating the couplings and the MVB masses. Using implicit
summation on repeated indices, the first of those is
\begin{align}
& f^5_{ajp}f^5_{dbp} - f^5_{abp}f^5_{djp} 
= \frac{m_j^2 + m_b^2 - m_c^2}{2m_jm_b}\, f_{dac}f_{cbj} 
\label{eq:timo-del-siglo-xx} % (20)
\\
& + \biggl(\frac{m_k^2 + m_j^2 - m_d^2}{m_jm_k} \,
\frac{m_k^2 + m_b^2 - m_a^2}{4m_bm_k} \,f_{djk} f_{abk}
- [a \otto d] \biggr),
\nn
\end{align}
if $m_b, m_j > 0$. The sum over~$c$ is over all gauge bosons and the
sum over~$k$ runs only over massive ones.

In particular, setting $j = a$ and $d = b \neq a$, one infers that
\begin{align}
& f^5_{aap} f^5_{bbp} - f^5_{abp} f^5_{bap}
\nn \\
&= \frac{1}{2m_a m_b} \biggl[\,\sum_{c=1}^t 
(m_a^2 + m_b^2 - m_c^2)\,(f_{abc})^2
\nn \\
&\qquad + \sum_{k:\,m_k\neq 0}\!
\frac{(m_a^2 - m_b^2)^2 - m_k^4}{2m_k^2} \,(f_{abk})^2 \biggr].
\label{eq:deus-ex-machina} % (21)
\end{align}
On the one hand this relation allows us to compute $f^5$ from the
masses and the structure constants; on the other hand, since it is
valid for any $b \neq a$, it implies direct relations between the
masses and the structure constants.%
\footnote{We observe that Scharf writes the previous equation
differently, since he mistakenly ``derived'' $f^5_{abp} = 0$ when
$a \neq b$.}

It will help to reorganize \eqref{eq:deus-ex-machina}, separating the
massless from the massive bosons in the sum. If $m_k \neq 0$, the
coefficient of $(f_{abk})^2 /4m_a m_b m_k^2$ is
\begin{align*}
& 2m_k^2(m_a^2 + m_b^2 - m_k^2) + (m_a^2 - m_b^2)^2 - m_k^4
\\
&\quad = (m_a^2 + m_b^2 + m_k^2)^2 - 4(m_a^2 m_b^2 + m_k^4).
\end{align*}
Thus the main consequence of \eqref{eq:timo-del-siglo-xx} can be
written, for $a \neq b$ with $m_b \neq 0$, as:
\begin{align}
& 4m_a m_b \sum_{p=1}^z \begin{vmatrix} f^5_{aap} & f^5_{abp} \\[\jot]
f^5_{bap} & f^5_{bbp} \end{vmatrix}
= 2(m_a^2 + m_b^2) \sum_{d:\,m_d=0}\! (f_{abd})^2
\nn \\
&+ \sum_{k:\,m_k\neq 0}\! \frac{(f_{abk})^2}{m_k^2}
\bigl[ (m_a^2 + m_b^2 + m_k^2)^2 - 4(m_a^2 m_b^2 + m_k^4) \bigr].
\label{eq:minas-gerais} % (22)
\end{align}
This first constraint and \eqref{eq:madre-del-cordero}, with their
respective consequences \eqref{eq:minas-gerais} and
\eqref{eq:the-thin-edge}, restrict strongly the masses of the gauge
bosons (subsections~\ref{ssc:SU2}--\ref{ssc:SUn}).

% \S 4
\section{The relation between CGI and SSB}
\label{sec:Sherlock}

The primary aim of this section is to work out explicitly the
connection of model building by CGI to the SSB approach. In particular
we show that one obtains the covariant derivative of the scalar
fields, that is, the ``minimal coupling'' recipe. A related aim is to
disprove the claim~\cite{FortunaJuvet} about standard GUT models not
satisfying CGI at tree level, that would contradict our aforementioned
statement. Finally, we collect information on the $L_2$ piece of the
Lagrangian. We restate that:
\begin{itemize}
\item
{}From CGI for spin-one particles, one is led to discover the gauge
symmetry: the coupling parameters $f_{abc}$ in~$L_1^1$ are the
structure constants of a reductive Lie algebra, and the other
couplings $f^*_{***}$ in~$L_1^*$ are determined by the~$f_{abc}$ and
the masses. Knowledge of this hidden symmetry is of course very
useful, but not needed \textit{a priori} within CGI.
\item
In the opposite direction, i.e. postulating the underlying gauge
symmetry, we expect that models built by SSB be classically BRST
invariant, and hence satisfy CGI at tree level to all
orders~\cite{Michael05}.
\end{itemize}

% \S 4.1
\subsection{Reinterpreting the first constraint from CGI}
\label{ssc:al-toro}

Using~\eqref{eq:little-lamb}, the main obstruction
\eqref{eq:timo-del-siglo-xx} is rewritten
\begin{align}
f^5_{ajp} f^5_{dbp} - f^5_{abp} f^5_{djp}
= 2 f_{dac} f^3_{cbj} + 4 f^3_{ajk} f^3_{dkb} - 4 f^3_{abk} f^3_{dkj}.
\label{eq:timo-del-siglo-xxi} % (23)
\end{align}
In view of \eqref{eq:higgfree} and~\eqref{eq:higgful} it is clear
that $f^3$ and $f^5$ should be related. We introduce the notation
$$
(F^a)_{bc} = -2 f^3_{abc}
$$
for $r \x r$ skewsymmetric matrices $F^a$. If we provisionally assume
that only one physical scalar is present $(z = 1)$, let $G^a$ be
$r \x 1$ matrices (there are $r + s$ of these) given by
$$
(G^a)_j = - f^5_{aj},
$$
and form the $(r + 1) \x (r + 1)$ skewsymmetric matrices,
for $a = 1, \dots, r + s$:
$$
S^a = \begin{pmatrix} F^a & G^a \\ -{}^tG^a & 0 \end{pmatrix},
\text{ with $^tG^a$ being the transpose of $G^a$}.
$$
Relation \eqref{eq:timo-del-siglo-xxi} corresponds to the left upper
corner of the commutator bracket
\begin{equation}
[S^a, S^d] = f_{adc}\, S^c.
\label{eq:addio} % (24)
\end{equation}
Employing $f^5_{ab} = C_5\,m_a\,\dl_{ab}$ and~\eqref{eq:any-wool}, one
sees that the other corners of this bracket formula are fulfilled,
too. Thus equation~\eqref{eq:timo-del-siglo-xxi} means that $f^3$,
$f^5$ taken together define a real skewsymmetric matrix representation
of the gauge group with dimensionless entries, for only one higgs.

% \S 4.2
\subsection{How the covariant derivative arises from CGI}
\label{ssc:a-la-vaca}

When more than one higgs is present, one should admit terms like
$(A_a \. \del\vf_q)\vf_p - (A_a \. \del\vf_p)\vf_q$, and so we have
done in~\eqref{eq:higgful}. In this case a second constraint is found,
$$
0 = f_{abc} f^8_{dcp} - f^4_{dbk} f^5_{akp} + f^4_{dak} f^5_{bkp}
+ 2 f^6_{apv} f^8_{dbv} - 2 f^6_{bpv} f^8_{dav}\,.
$$
Here $a,b,d,p$ are fixed; the summation indices are $c = 1,\dots,r+s$;
$k = 1,\dots,r$; and $v = 1,\dots,z$. The right hand side is the
coefficient of the term $[u_a u_b \ut_d \vf_p](x)\,\dl(x - y)$ on the
right hand side of the CGI condition~\eqref{eq:CGI-2} for $n = 2$. We
point out that the expression must be antisymmetric in $a \otto b$
because $u_a u_b \ut_d \vf_p$ is. There is no contribution coming from
$[Q_\mathrm{in},\ N_2]$ ---hence the zero on the left hand side---
since quartic terms involving ghost fields $u$, $\ut$ are not admitted
here. This can be justified by the second CGI method.

\smallskip

We finally line up the following system of constraints:
\begin{align}
& 4 f^3_{adk} f^3_{bke} - 4 f^3_{bdk} f^3_{ake}
- f^5_{adv} f^5_{bev} + f^5_{bdv} f^5_{aev}
\nn \\
&\quad = - 2 f_{abc} f^3_{cde}\,,  \quad (m_d > 0,\ m_e > 0) 
\nn \\[2\jot]
& 2 f^3_{adk} f^5_{bkp} - 2 f^3_{bdk} f^5_{akp}
- 2 f^5_{adv} f^6_{bpv} + 2 f^5_{bdv} f^6_{apv} 
\label{eq:mater-et-magistra} % (25)
\\
&\quad = - f_{abc} f^5_{cdp}\,,    \quad\,\ (m_d > 0)  
\nn \\[2\jot]
& - f^5_{akp} f^5_{bkq} + f^5_{bkp} f^5_{akq}
+ 4 f^6_{apv} f^6_{bvq} - 4 f^6_{bpv} f^6_{avq}
\nn \\
&\quad = - 2 f_{abc} f^6_{cpq}\,.
\nn
\end{align}
The first equation is the by now familiar basic constraint of Sect.\
\ref{ssc:ties-that-bind}; the second is the previously displayed
equation divided by~$-m_d$. The derivation of these constraints from
CGI is discussed in Appendix \ref{app:constraints}.

Let us reintroduce the matrices $(G^a)_{dp} = - f^5_{adp}$, which are
now $r \x z$, and introduce the $z \x z$ ones:
$$
(H^a)_{pq} = -2 f^6_{apq}.
$$
Then the system \eqref{eq:mater-et-magistra} amounts to the triplet
of matrix equations,
\begin{align*}
[F^a, F^b] - G^a\,{}^tG^b + G^b\,{}^tG^a &= f_{abc}\, F^c,
\qquad  (r\x r)
\\[2\jot]
F^a G^b - F^b G^a + G^a H^b - G^b H^a &= f_{abc}\, G^c,
\qquad  (r\x z)
\\[2\jot]
- {}^tG^a G^b + {}^tG^b G^a + [H^a, H^b] &= f_{abc}\, H^c.
\qquad  (z\x z).
\end{align*}
Putting it all together in the skewsymmetric $(r+z) \x (r+z)$ package
$$
S^a = \begin{pmatrix} F^a & G^a \\ -{}^tG^a & H^a \end{pmatrix}
= \begin{pmatrix} -2 f^3_{a**} & - f^5_{a*\star} \\
f^5_{a\star*} & -2 f^6_{a\star\star} \end{pmatrix},
$$
this generalizes the \textit{gauge-group representation}
$[S^a,S^b] = f_{abc}\,S^c$ mooted in~\eqref{eq:addio}. In fine, the
key couplings $f^5$, $f^6$ of the physical scalars are constrained by
this algebraic relation in terms of the ``known'' $f^3$ couplings.%
\footnote{Also within CGI, in an analogous way the coupling of vector
bosons to fermions induces a gauge-group representation among the
latter fields.}

Moreover, we contend that the representation above is the one yielding
the covariant derivative on the scalar multiplets of the ``minimal
coupling'' recipe, written in real form. More explicitly, let $\eta$
be a scalar multiplet assembled from the St\"uckelberg fields and the
higgses by
\begin{equation}
\eta^t := (B_1, \dots, B_r, \vf_1 + v_1, \dots, \vf_z + v_z),
\label{eq:eta} % (26)
\end{equation}
where the field shifts $v_p$ are real numbers. We make two assertions.
The first is a statement about the corresponding hidden gauge
symmetry; namely that the multiplet $\eta$ transforms with the
representation $S^a =: S(T_a)$, with $T_a$ the generators of the gauge
Lie algebra, and hence
$$
D^\mu\eta: = (\del^\mu + \ka A^\mu_a S^a)\,\eta,
$$
is the covariant derivative of~$\eta$. Our second claim is that the
minimal coupling recipe holds true, in the sense that, with a suitable
choice of the~$v_p$,
\begin{align}
\frac{1}{2} \del\eta^t \. \del\eta
&+ \frac{\ka}{2} \bigl( (A_a \. \del\eta^t\,) S^a\eta
- \eta^t S^a(A_a \. \del\eta) \bigr)
\nn \\
&\quad - \frac{\ka^2}{4} (A_a \. A_b) \eta^t\,[S_a, S_b]_+ \eta
\label{eq:mini-couple} % (27)
\end{align}
agrees with what one obtains by the CGI method for the scalar-gauge
Lagrangian, that is, besides the kinetic terms of the $B$- and
$\vf$-fields, the vector-boson mass term, plus an $(A\.\del B)$ term,
plus the trilinear couplings $L_1^2 + L_1^3 + L_1^5 + L_1^6 + L_1^7$
of the gauge fields $A$ to the scalars $(B,\vf)$ considered in
\eqref{eq:higgfree} and~\eqref{eq:higgful}, plus the quartic terms
$L_2^2 + L_2^3 + L_2^4$ defined in Sect.~\ref{ssc:look-farther} right
below. (Strictly speaking, within CGI the shift of the fields by the
$v_p$ is not required at second order. However, it is convenient for
our purposes. In examples, the ``correct'' choice of $v_p$ can be
obtained from a comparison with SSB: the fields $\vf_p + v_p$ are the
ones of the ``unbroken'' model with its full gauge symmetry.) We
routinely verify our assertions in the example models constructed by
CGI in the next section. Therefore our procedure derives within CGI
the crucial piece $\half(D\eta)^t \. D\eta$ of the Lagrangian.
Indeed this provides the crowning point of the construction.

% \S 4.3
\subsection{On the quartic couplings}
\label{ssc:look-farther}

There are quartic terms
$$
L_2 = \half(L_2^1 + L_2^2 + L_2^3 + L_2^4 + L_2^5 + L_2^6 + L_2^7),
$$
with obvious symmetries as before, of the following form:
\begin{align}
L_2^1 &= h^1_{bcde} \,(A_b \. A_d) (A_c \. A_e),
& L_2^5 &= h^5_{abcd} \,B_a B_b B_c B_d,
\nn \\
L_2^2 &= h^2_{abcd} \,(A_a \. A_b) B_c B_d,
& L_2^6 &= h^6_{abpq} \,B_a B_b \vf_p \vf_q,
\nn \\
L_2^3 &= h^3_{abcp} \,(A_a \. A_b) B_c \vf_p,
& L_2^7 &= h^7_{pqrs} \,\vf_p \vf_q \vf_r \vf_s.
\nn \\
L_2^4 &= h^4_{abpq} \,(A_a \. A_b) \vf_p \vf_q,
\label{eq:imperitia-culpae-adnumeratur} % (28)
\end{align}
A complete account of the permitted quartic terms would take us too
far afield. For instance, to answer the question of whether models are
completely fixed in the general case by CGI and by requiring that the
number of higgs fields be as small as possible, one needs a complete
study of the tree-level third-order conditions, as well as to revisit
some corners of the second-order conditions, here unexplored. This is
better left for another paper. We limit ourselves to reporting on what
can be gleaned from the foregoing and calculations analogous to the
ones performed in the coming Sect.~\ref{sec:CGI-works} and in
Appendix~\ref{app:constraints}.

One finds from CGI $h^1_{bcde} = -\half f_{abc}f_{ade}$ as thoroughly
expected: it just yields the quartic part in the Yang--Mills
Lagrangian, irrespectively of masses.

Now, it is plain what $L_2^2$, $L_2^3$, $L_2^4$ of formula
\eqref{eq:imperitia-culpae-adnumeratur} must be. Have a look back
at~\eqref{eq:mini-couple}. According to our results on minimal
coupling from CGI at second order, these terms in the interaction
Lagrangian are generated by suitable combinations \textit{not
involving} $v$ in $-\quarter(A_a\. A_b)\eta^t\,[S_a, S_b]_+\eta$.
Therefore, taking into account the factor $\half$ in the definitions,
one finds:
\begin{itemize}
\item
For the higgs-free term $L_2^2$,
$$
h^2_{abcd} = -2 f^3_{ack} f^3_{bkd} - 2 f^3_{bck} f^3_{akd}
+ \half f^5_{acv} f^5_{bdv} + \half f^5_{bcv} f^5_{adv}.
$$
Here and in the subsequent formulas we sum over repeated indices. This
is symmetric under $a \otto b$ and $c \otto d$, as it should be.

\item
$h^3_{abcp} = -2 f^3_{ack} f^5_{bkp} - 2f^3_{bck} f^5_{akp}
+ 2 f^5_{acv} f^6_{bpv} + 2 f^5_{bcv} f^6_{apv}$. This is symmetric in
$a,b$.

\item
$h^4_{abpq} = -2 f^6_{apv} f^6_{bvq} - 2 f^6_{bpv} f^6_{avq}
+ \half f^5_{akp} f^5_{bkq} + \half f^5_{bkp} f^5_{akq}$. This is 
symmetric under $a \otto b$ and $p \otto q$.
\end{itemize}

For the higgs-free term $L_2^5$ we find, for $a,b,c,d\leq r$:
$$
h^5_{abcd} = - \frac{\mu_p^2}{12} \biggl(
\frac{f^5_{abp} f^5_{cdp}}{m_am_c} + \frac{f^5_{acp} f^5_{bdp}}{m_am_b}
+ \frac{f^5_{adp} f^5_{cbp}}{m_am_c} \biggr).
$$
This ought to be symmetric under exchanges of $a,b,c,d$, and indeed it
is: the relations $m_b f^5_{abp} = m_af^5_{bap}$ save the day. For
$L_2^6$ we find:
\begin{align*}
h^6_{abpq} &= \frac{\mu_p^2 + \mu_q^2}{4m_a m_b}
(f^5_{acp} f^5_{bcq} + f^5_{acq} f^5_{bcp})
+ \frac{3}{m_a} f^{11}_{pqu} f^5_{abu}
\\ 
&\quad + \frac{2\mu_u^2 - \mu_q^2 - \mu_p^2}{m_a m_b}
(f^6_{aup} f^6_{buq} + f^6_{auq} f^6_{bup})
\\
&\quad + \frac{(m_a^2 - m_b^2)(\mu_p^2 - \mu_q^2)}{2m_a m_b m_k^2}
f_{abk} f^6_{kpq}.
\end{align*}
This has the required symmetries under $a\otto b$, $p\otto q$; it is
undetermined at second order, because $f^{11}$~is.

Finally, $h^7$ is undetermined at second order. CGI for third-order
tree diagrams yields conditions restricting $h^7$ and $f^{11}$ (via
conditions on $h^6$), which in the case $z=1$ determine these
parameters uniquely; see the next subsection. The procedure was
explained in~\cite[Sect.~5]{PGI-EW-II}, with calculations given in
detail for the SM; consult~\cite{Scharf} as well.

% \S 4.4
\subsection{Quartic couplings for models with only one higgs}
\label{ssc:look-again}

For the case $z = 1$, with the $\vf$-index suppressed, we obtain:
\begin{enumerate}
\item
\setlength{\abovedisplayskip}{-14pt}
\begin{align*}
h^2_{abcd} = &\sum_{k:\,m_k\neq 0} \frac{1}{8m_c m_d m_k^2}
\bigl( f_{ack} f_{bdk}
\\
&\x (m_c^2 + m_k^2 - m_a^2)(m_d^2 + m_k^2 - m_b^2)
 + [a \otto b] \bigr)
\\
&+ \half\,C_5^2\,m_a m_b (\dl_{ac}\,\dl_{bd} + \dl_{ad}\,\dl_{bc}).
\end{align*}

\item
$h^3_{abc} = 2f^2_{abc} C_5 = f_{abc} C_5 (m_b^2 - m_a^2) /m_c$.

\item
$h^4_{ab} = C_5^2 m_a^2\,\dl_{ab}$. 

\item
$h^5_{abcd} = \third (\dl_{ab}\,\dl_{cd} + \dl_{ac}\,\dl_{bd}
+ \dl_{ad}\,\dl_{bc})\,h^7\,$; $h_{ab}^6 = 2\dl_{ab}\,h^7\,$;
$h_7 = -\quarter C_5^2 m_H^2$, independently of indices~$\leq r$.
\end{enumerate}

This allows us to peek at the purely scalar sector with one higgs.
Including its mass term, it becomes
\begin{align}
& -\frac{1}{2} m_H^2 \vf^2 + \ka(f^9_{ab} B_a B_b \vf + f^{11} \vf^3)
\nn \\
&\quad + \frac{\ka^2}{2} (h^5_{abcd} B_a B_b B_c B_d
+ h^6_{ab} B_a B_b \vf^2 + h^7 \vf^4)
\nn \\
&= -\frac{m_H^2}{2} \biggl( \vf^2 
+ \ka C_5 \Bigl(\, \sum_{a=1}^r B_a^2 + \vf^2 \Bigr) \vf
\nn \\
&\quad + \frac{\ka^2}{4} C^2_5 \Bigl(\,
\sum_{a=1}^r B_a^2 + \vf^2 \Bigr)^2 \biggr)
\nn \\
&= -\frac{\ka^2 m_H^2 C_5^2}{8} \biggl(\frac{2\vf}{\ka C_5} + \vf^2
+ |\vec{B}|^2\biggr)^2 =: -V(\vf,\vec{B}).
\label{eq:hoc-erat-in-votis} % (29)
\end{align}
These formulas are correctly given in~\cite{Scharf}. The potential
exhibits a characteristic $O(r+1)$ symmetry~\cite{SmokeScreen}.
Leaving aside the St\"uckelberg fields, it has a minimum at 
$\vf = 0$. Hence, the physical higgs field can be realized in an
ordinary Fock representation, with a unique vacuum and
\textit{vanishing} vacuum expectation value.

% \S 5
\section{The CGI methods in practice}
\label{sec:CGI-works}

The plan of this section is as follows. We first attack from the
perspective of the first CGI approach the simplest example one can
think of ---dealt with only summarily in~\cite{Scharf}. We investigate
next models with several massive vector bosons, but one physical higgs
($z = 1$) only, using the first CGI method as in
Sect.~\ref{ssc:ties-that-bind}. One may derive here the
\textit{possible mass patterns} of the gauge bosons by taking only the
consequences of equations \eqref{eq:madre-del-cordero} and
\eqref{eq:timo-del-siglo-xx} into account. Of course, to show that the
resulting models indeed satisfy CGI at tree level, one must verify
\textit{all} $c$-number identities expressing \eqref{eq:CGI1} and
\eqref{eq:CGIn} on that level. The solutions of those equations that
we work out are compatible with the CGI conditions at all orders. We
finally look at causal gauge invariance for models with scalar fields
in the adjoint. All along, we flesh out the relation between CGI and
SSB whose theoretical underpinning was derived in the previous
section.

% \S 5.1
\subsection{The toy model}
\label{ssc:Abel}

The case $r = 1$, $s = 0$, $z = 1$ leads to an Abelian model in which
all the terms $L_1^5$ to~$L_1^{11}$ with the higgs-like field $\vf$
appear, except~$L_1^6$. All contributions of the first group, $L_1^1$
to~$L_1^4$, disappear. Also $L_2^1$ and $L_2^2$ vanish. The
obstructions of Sect.~\ref{ssc:ties-that-bind} play no role here.
This does not sound very interesting; but it is instructive. Eleven
contributions in all survive, we find that $C_5 = 1/m$ with $m$ being
the mass of the spin~1 particle, and the resulting interaction
Lagrangian reads
\begin{align}
L_\mathrm{int}(x) 
&= \ka m(A\.A)\vf - \ka m\ut u\vf + \ka B(A\.\del\vf)
\nn \\
&\quad - \ka \vf(A\.\del B) - \frac{\ka m_H^2}{2m} \vf^3
- \frac{\ka m_H^2}{2m} B^2\vf
\nn \\[\jot]
&\quad + \frac{\ka^2}{2} (A\.A) \vf^2 + \frac{\ka^2}{2} (A\.A) B^2
- \frac{\ka^2m_H^2}{8m^2} \vf^4 
\nn \\
&\quad - \frac{\ka^2m_H^2}{4m^2} \vf^2 B^2
- \frac{\ka^2m_H^2}{8m^2} B^4,
\label{eq:anatema-sit} % (30)
\end{align}
where $m_H$ is the mass of the higgs field~$\vf$.

For the derivation of~\eqref{eq:anatema-sit}, recall that $T_1 = L_1$
is given by the first two lines of~\eqref{eq:anatema-sit}. Assume that
CGI to first order \eqref{eq:CGI1} has already been put to work,
yielding the first six terms on the right hand side
in~\eqref{eq:anatema-sit}, except that the coefficient of the
$\vf^3$-coupling is undetermined. The $Q$-vertex here is given by
$$
s_0 T_1 = \del\. P  \text{ with }
P = \ka\bigl( mu\vf A - u(\vf\,\del B - B\,\del\vf) \bigr).
$$
Next put to work CGI for second-order tree diagrams~\eqref{eq:CGI-2}.
As a rule, calculations of this kind are elementary, but tedious.
Unhurried readers are referred to the leisurely treatment
in~\cite{Ausonia}. Also, a technically more detailed version of this
paper, containing the computations pertaining here in particular, is
available as~hep-th/1001.0932v2.

% \S 5.1.1
\subsubsection{The second CGI criterion and gauge independence}
\label{sss:gauge-indep}

For the sake of training, we wish to verify the second CGI criterion
in this example directly. So far we have adhered to the Feynman gauge,
whereby the masses of gauge and St\"uckelberg fields coincide. To show
that this restriction is not necessary, we proceed here in an
arbitrary $\La$-gauge \textit{\`a la} 't~Hooft.

It is instructive to look first at the \textit{free} model. The
St\"uckelberg Lagrangian for a MVB is most elegantly written
\cite{Altabonazo,Felicitas}:
$$
L_\mathrm{Stue} 
= L_\kin(A) + \frac{m^2}{2} \biggl( A - \frac{\del B}{m} \biggr)^{\!2}
- \frac{\La}{2} \biggl( \del\. A + \frac{m}{\La} B \biggr)^{\!2},
$$
where $\La$ is the gauge-fixing parameter. The first two terms are
manifestly gauge invariant by $\dl A = \del\al$, $\dl B = m\al$;
however, the last one (which is the gauge-fixing term $L^\gf_0$) is
gauge invariant only if $(\square + m^2/\La)\al = 0$.

For the zeroth order BRST transformation, the gauge-fixing parameter
$\La$ appears only in $s_0 \ut$:
$$
s_0 \ut = -(\La\,\del\.A + mB)
$$
and $s_0 A = \del u$, $s_0 B = mu$, $s_0 u = 0$, $s_0\vf = 0$ as in
the Feynman gauge. Obviously $s_0$ is nilpotent, except maybe for
$s_0^2\ut$; we return to this point below.

The first two terms in $L_\mathrm{Stue}$ are $s_0$-invariant, but for
an unrestricted $u$-field this does not hold for~$L^\gf_0$:
\begin{align*} 
s_0 L^\gf_0 
&= \bigl( \del\. s_0 A + \tfrac{m}{\La}\, s_0 B \bigr)\, s_0 \ut
= \bigl( \square + \tfrac{m^2}{\La} \bigr) u\, s_0 \ut.
\end{align*} 
For this reason one introduces a ghost Lagrangian $L^\gh_0$ such that
$s_0(L^\gf_0 + L^\gh_0)$ is a divergence: with
\begin{align}
L^\gh_0 &= \del\ut\.\del u - \frac{m^2}{\La}\, \ut u
= \del\ut\. s_0 A - \frac{m}{\La}\, \ut s_0 B,
\label{eq:ghosty} % (31)
\end{align}
we indeed obtain
$$
s_0 (L^\gf_0 + L^\gh_0) = \del\.(s_0 \ut\, s_0 A) =: \del\. I_0.
$$
Adding $L^\gh_0$ and the kinetic and mass terms for the higgs to
$L_\mathrm{Stue}$, the total free Lagrangian $L_0$ takes the form
\begin{align*} 
L_0 &= L_\kin(A) + \frac{m^2}{2} (A\. A) 
+ \frac{1}{2} (\del B\. \del B) - \frac{m^2}{2\La}\, B^2 
\\[\jot] 
&\quad - \frac{\La}{2} (\del A)^2 
+ \frac{1}{2} (\del\vf\. \del\vf) - \frac{m_H^2}{2}\, \vf^2 
+ \del\ut\. \del u 
\\
&\quad - \frac{m^2}{\La}\, \ut u - m\,\del\.(AB). 
\end{align*} 
It is BRST invariant in the sense that $s_0L_0 =  \del\. I_0$.

Returning to nilpotence of $s_0$, we see that $s_0^2\ut$ vanishes
modulo the free field equations:
$$
s_0^2\ut = -\La\,\del(s_0 A) - m s_0B = -\La\square u - m^2u
= \La\,\frac{\dl S_0}{\dl\ut},
$$
where $S_0$ is the action corresponding to~$L_0$. The equations of
motion for the free vector field $A$ and the St\"uckelberg field~$B$
are seen to be
$$
(\square + m^2) A = (1 - \La)\,\del(\del\.A); \qquad
(\square + \La^{-1}m^2) B = 0.
$$
Thus, if any other than the Feynman gauge $\La = 1$ is chosen, the
mass of the St\"uckelberg field becomes $m/\sqrt{\La}$; this is also
the mass of the ghost fields $u$, $\ut$, and of~$\del A$.

\medskip

Turning to the interacting sector, we need to verify BRST invariance
for the terms \eqref{eq:anatema-sit} obtained by the first CGI method.
The interacting BRST transformation $s = s_0 + s_1$ has an additional
term $s_1 \sim \ka$, given by
\begin{equation}
s_1 B = \ka\, u\vf,  \quad  s_1 \vf = - \ka\, Bu,
\label{eq:sapiens-nihil-afirmat} % (32)
\end{equation}
and zero for the other fields.
Let us look immediately at the scalar sector. In this instance
$\vec{B}$ of~\eqref{eq:hoc-erat-in-votis} has a single component, and
the point is that \eqref{eq:sapiens-nihil-afirmat} guarantees that
$$
sV(\vf,B) \propto
(s_0 + s_1) \biggl( \frac{2m\vf}{\ka} + \vf^2 + B^2 \biggr)^2 = 0.
$$
{}From $su = 0$ and $uu = 0$ we obtain $s^2 A = s^2 B = s^2 \vf = 0$.
For $L^\gh$ we keep the form $L^\gh = \del\ut\. sA -
\tfrac{m}{\La}\,\ut sB$ in~\eqref{eq:ghosty}. Note that this
contributes the second term in~\eqref{eq:anatema-sit}, when $\La = 1$.
Still with the new action $S$, we find that
$$
s^2 \ut = -\La\,\del(s A) - m sB = \La\,\frac{\dl S}{\dl\ut}
$$
vanishes on-shell, since only $L^\gh$ contributes to $\dl S/\dl\ut$.
The gauge-fixing part is not modified, and again
\begin{align*}
s(L^\gf + L^\gh)
&= (\del\. sA + \tfrac{m}{\La}\, sB) s\ut + \del(s\ut)\.sA
- \tfrac{m}{\La} s\ut\, sB
\\
&= \del\.(s\ut\, sA) = \del\. I_0 \,.
\end{align*}
In this particularly simple case, the vector $I$ has only components
of degree zero.

We know that $s L_\kin(A) = 0$. The total Lagrangian reads
\begin{align*}
L &= L_\kin(A) + \frac{m^2}{2} \Bigl(A - \frac{\del B}{m} \Bigr)^2
+ \frac{1}{2} \,\del\vf\.\del\vf - \frac{m_H^2}{2} \vf^2
\\
&\quad + L^\gf + L^\gh + (L_\mathrm{int} + \ka m \ut u \vf),
\end{align*}
where $L_\mathrm{int}$ is given by~\eqref{eq:anatema-sit}. It remains
to verify BRST invariance of
$$
L - L_\kin(A) - L^\gf - L^\gh + V(\vf,B) =: L^\eta.
$$
As discussed in Sect.~\ref{sec:Sherlock}, these terms can be grouped
into a minimal coupling recipe. As an example of the
$S$-represen\-tation of that section, we have the sole $S$-matrix
$\begin{pmatrix} 0 & -1 \\ 1 & 0 \end{pmatrix}$. Let us use it to the
purpose. With $\eta = (B, m/\ka + \vf)^t$ and $D = \del + \ka\,A\,S$,
we obtain
\begin{align*}
&\half (D\eta)^t \. D\eta 
\\
&\quad = \frac{1}{2}
\begin{pmatrix} \del B - \ka(m/\ka + \vf)A \\[\jot]
\del\vf + \ka BA \end{pmatrix}^t \begin{pmatrix}
\del B - \ka(m/\ka + \vf)A \\[\jot] \del\vf + \ka BA \end{pmatrix}
\\
&\quad = \half\,\del B\.\del B + \half\,\del\vf\. \del\vf
+ \half m^2 A\.A - mA\.\del B 
\\
&\qquad + \ka\bigl( m(A\.A)\vf + B(A\.\del\vf) - \vf(A\.\del B) \bigr)
\\
&\qquad + \half\ka^2 \bigl( (A\. A)\vf^2 + (A\. A)B^2 \bigr) = L^\eta,
\end{align*}
indeed providing the sought-for terms. Since the BRST variation
of~$\eta$ has the form of an infinitesimal gauge transformation,
$$
s\eta = \begin{pmatrix} sB \\ s\vf \end{pmatrix} 
= \begin{pmatrix} u(m + \ka\vf) \\ -\ka uB \end{pmatrix}
= -\ka u\, S\eta,
$$
the covariant derivative satisfies $s\,D\eta = -\ka u\,SD\eta$, and
hence $sL^\eta = 0$. We conclude that our toy model is BRST invariant,
thus causal gauge invariant on the tree level, and that the
first~\cite{Scharf} and second~\cite{Michael05} CGI criteria match for
it.

% \S 5.1.2
\subsubsection{Comparison with SSB}
\label{sss:match-SSB}

The model with one massive and no massless gauge boson we have been
working with can obviously be obtained by SSB of an $U(1)\simeq O(2)$
gauge model. Let us employ instead of~$\eta$ the complex field
$\Phi := iB + m/\ka + \vf$. The real part of~$\Phi$ is interpreted as
a shifted higgs-like field $H = 1/\ka C_5 + \vf = m/\ka + \vf$, and we
rewrite \eqref{eq:hoc-erat-in-votis} in terms of it, obtaining the
quartic polynomial
$$
V(\Phi) = V_0 - \frac{\mu^2}{2}\, \Phi^t \Phi
+ \frac{\la}{4} (\Phi^t \,\Phi)^2 =: V_0 + V_\mathrm{mod}(\Phi),
$$
where
$$
V_0 = \frac{m_H^2 m^2}{8\ka^2}; \quad \mu = \frac{m_H}{\sqrt{2}};
\quad \la = \frac{\ka^2 m_H^2}{2m^2}.
$$
In order to extract the SSB model from this, drop the constant term
---this has ``only'' epistemological and gravity-cosmological
consequences~\cite{Tini75}. Then seek the minimum of the potential
$\dl V/\dl\Phi = (-\mu^2+\la\, \Phi^t\,\Phi)\,\Phi = 0$. Any solution
of this can be ``rotated'' to a real value
$\val{\Phi} = \mu/\sqrt{\la} = m/\ka =: v$. Patently we have
reconstructed the ``Abelian Higgs model'', in which an initially
massless vector boson~$A$ is held to acquire the mass $m = \ka v$. (A
pity that we cannot switch off the interaction to see whether $A$ was
indeed massless.) The remaining scalar particle $\vf$, or higgs,
corresponding to the perturbation of $\Phi$ with respect to~$v$, has a
mass $\sqrt{2\la}\,v$, which is precisely~$m_H$.

% \S 5.2
\subsection{$SU(2)$ models with only one higgs}
\label{ssc:SU2}

With three gauge fields, the only relevant Lie algebra entering the
game is $SU(2)$; this means we take $f_{abc} = \eps_{abc}$, whereupon
total antisymmetry implies the Jacobi identity. This is surely an
important case.
\begin{enumerate}
\item
The case $m_1 = m_2 = m_3 = 0$ is certainly possible, and then neither
higgses nor St\"uckelberg fields are necessary.

\item
We see from~\eqref{eq:madre-del-cordero} that if $m_3 = 0$ then
$m_1 = m_2$ must hold; the pattern $m_2 = m_3 = 0$, $m_1 \neq 0$ is
downright forbidden.

\item
The case $m_3 = 0$, $m_1 = m_2 \neq 0$, after the necessary checks of
all CGI tree-level conditions, turns out to be possible with one
higgs-like field.

\item
Finally, if we assume that all masses are different from zero, then
necessarily $m_1 = m_2 = m_3$. This last case, also after all
necessary checks, turns out to be possible as well with one higgs-like
field.
\end{enumerate}

Physically, the two cases just mentioned correspond respectively to
the Georgi and Glashow ``electroweak'' theory without neutral
currents; and to the $SU(2)$ Higgs--Kibble model. Both can be thought
of as a limit of the SM, in the second case by setting the Weinberg
angle equal to zero and dropping the decoupled photon field.

With respect to the pending checks, let us show first why
$m_1 = m_2 = m_3$ must hold when there are three massive gauge fields.
Indeed, equation~\eqref{eq:minas-gerais} implies
$$
4 m_a^2 m_b^2 m_c^2 C_5^2 = (\eps_{abc})^2 \bigl[
(m_a^2 + m_b^2 + m_c^2)^2 - 4(m_a^2 m_b^2 + m_c^4) \bigr],
$$
where $(a,b,c)$ is any permutation of $(1,2,3)$. Therefore,
$$
m_1^2 m_2^2 + m_3^4 = m_2^2 m_3^2 + m_1^4 = m_3^2 m_1^2 + m_2^4.
$$
This yields
\begin{align*}
& (m_1^2 m_2^2 + m_3^4) - (m_2^2 m_3^2 + m_1^4)
\\
&\qquad = (m_3^2 - m_1^2)(m_1^2 - m_2^2 + m_3^2) = 0,
\nn \\
& (m_2^2 m_3^2 + m_1^4) - (m_3^2 m_1^2 + m_2^4)
\\
&\qquad = (m_1^2 - m_2^2)(m_2^2 - m_3^2 + m_1^2) = 0,
\end{align*}
whose only all-positive solution is $m_1 = m_2 = m_3 =: m$; and then
$4m^6 C_5^2 = m^4$ imposes $C_5 = 1/2m$. Formula
\eqref{eq:the-thin-edge} is void here; the
test~\eqref{eq:timo-del-siglo-xx} is cleared as well. This model has
been exhaustively studied in \cite{CabezondelaSal} and~\cite{MWI}.
Note that actually $L_1^2 = 0 = L_2^3$ for~it.

For the other MVB model with $m_3 = 0$, equation
\eqref{eq:the-thin-edge} is clearly verified for all values of
$(a,b)$. As noted earlier, the equality $m_1 = m_2$ can already be
deduced from \eqref{eq:madre-del-cordero}, or from
\eqref{eq:the-thin-edge} alone. Now we find $C_5 = 1/m$.

In both $SU(2)$ cases with massive gauge bosons, the couplings are
completely determined from CGI, without SSB playing any role. It is
nevertheless quite easy to identify the corresponding models in the
framework of the Higgs mechanism, with the help of the $S$-matrices.
For the three-boson model with one vanishing mass, the results for
$f^3$ and $f^5$ here give
$$
S^1 = \begin{pmatrix} 0 & 0 & -1 \\ 0 & 0 & 0 \\ 1 & 0 & 0
\end{pmatrix}; \ 
S^2 = \begin{pmatrix} 0 & 0 & 0 \\ 0 & 0 & -1 \\ 0 & 1 & 0
\end{pmatrix}; \ 
S^3 = \begin{pmatrix} 0 & -1 & 0 \\ 1 & 0 & 0 \\ 0 & 0 & 0
\end{pmatrix},
$$
and clearly~\eqref{eq:addio} holds. For the three-boson model with
three equal masses, suppressing some null entries in the notation, we
likewise get
\begin{gather*}
\renewcommand{\arraycolsep}{3pt} %% narrower matrix columns here
S^1 = \begin{pmatrix} && 0 & -\half \\ && -\half & 0 \\
0 & \half \\ \half & 0 \end{pmatrix};  \quad
S^2 = \begin{pmatrix} && \half & 0 \\ && 0 & -\half \\ 
-\half & 0 \\ 0 & \half \end{pmatrix};
\\
S^3 = \begin{pmatrix} 0 & -\half \\ \half & 0 \\ 
&& 0 & -\half \\ && \half & 0 \end{pmatrix}.
\end{gather*}

%\S 5.3
\subsection{$SU(n)$ models with only one higgs}
\label{ssc:SUn}

For $SU(3)$ take the basis of Gell-Mann matrices $T_a = \la_a/2$, for
$a = 1,\dots,8$, normalized by $\tr(T_a T_b) = \half\,\dl_{ab}$. The
well-known structure constants, defined by
$[T_a,T_b] = i\,f_{abc}\,T_c$, are
\begin{gather*}
f_{123} = 1;  \quad
f_{147} = f_{246} = f_{257} = f_{345} = \half;
\\
f_{156} = f_{367} = -\half;   \quad
f_{458} = f_{678} = \tfrac{\sqrt{3}}{2},
\end{gather*}
and $f_{abc} = 0$ in all cases not arising from these by permuting
indices.

It is instructive to play with different mass patterns.

(i) Does the set of constraints allow an $SU(3)$ model with
\textit{all} masses positive? If $a \neq b$ and $f_{abk} \neq 0$ for
exactly one value of~$k$, then~\eqref{eq:minas-gerais} simplifies to
$$
4 m_a^2 m_b^2 m_k^2 C_5^2 = (f_{abk})^2 \bigl[ 
(m_a^2 + m_b^2 + m_k^2)^2 - 4(m_a^2 m_b^2 + m_k^4) \bigr],
$$
and, just as in Sect.~\ref{ssc:SU2}, invariance of
$m_a^2 m_b^2 + m_k^4$ under permutations of $a,b,k$ implies that
$m_a = m_b = m_k$. Applying this procedure for $(a,b,k) = (1,2,3)$,
$(1,4,7)$, $(2,4,6)$ and $(2,5,7)$ shows that
$m_1 = m_2 = m_3 = m_4 = m_5 = m_6 = m_7$. However, it should be noted
that the cases $(a,b) = (1,2)$ and $(1,4)$ respectively lead to
\begin{align*}
4 m_1^4 (f^5_{11})^2 &= (f_{123})^2 \,m_1^4 = m_1^4,
\\
4 m_1^4 (f^5_{11})^2 &= (f_{147})^2 \,m_1^4 = \quarter m_1^4.
\end{align*}
Therefore the inequality $f_{123} \neq \pm f_{147}$ yields an
impossibility: there is no all-massive $SU(3)$ model within our
approach. Note that $SU(2)$ escapes this sentence because all squared
structure constants are equal. The reader should be able to check that
the same phenomenon raises obstructions to several other putative
$SU(3)$ models.

(ii) This leads us to ponder the ``natural'' pattern:
$$
m_1 = m_2 = m_3 = 0;  \qquad  m_4, m_5, m_6, m_7 \neq 0.
$$
Indeed, the ``photons'' $m_1$, $m_2$ force
$m_4 = m_5 = m_6 = m_7 =: m \neq 0$ through use
of~\eqref{eq:minas-gerais}. Then $f^5_{44} = \half$ by just
considering in this equation $(a,b) = (4,6)$, say. By considering
$(a,b) = (4,5)$, one obtains $m_8 = 2m/\sqrt{3}$, and after some work,
it is checked that there is no contradiction in this. Note that
$(a,b,k) = (4,5,8)$ is not symmetrical with $(a,b,k) = (4,8,5)$, since
in one case there is a massless contribution ($f_{453} = \half$), but
not in the other. In conclusion: the model
\begin{gather}
m_1 = m_2 = m_3 = 0; \quad m_4 = m_5 = m_6 = m_7 = m;
\nn \\
m_8 = \frac{2}{\sqrt{3}}\,m \neq 0
\label{eq:dura-lex} % (33)
\end{gather}
solves the CGI mass conditions \eqref{eq:madre-del-cordero}
and~\eqref{eq:timo-del-siglo-xx}.

\smallskip

Turning to $SU(4)$, we can take basis matrices $\{T_a\}$ extending
those of $SU(3)$, filled out with a fourth row and column of zeroes,
by $\{T_9,\dots,T_{15}\}$, where
$$
T_{15} = \frac{1}{2\sqrt 6} \diag(1, 1, 1, -3),
$$
so that $\{T_3,T_8,T_{15}\}$ spans the Cartan subalgebra of diagonal
matrices, and the off-diagonal ones are the hermitian matrices given
in terms of the matrix units $e_{ij}$ by%
\footnote{In the standard notation for root vectors, the simple roots
$\al,\bt,\ga$ give $E_\al = e_{12}$, $E_\bt = e_{23}$,
$E_\ga = e_{34}$, $E_{\al+\bt} = e_{13}$, $E_{\bt+\ga} = e_{24}$,
$E_{\al+\bt+\ga} = e_{14}$. Note that $\al + \ga$ is not a root of
$SU(4)$ since $[E_\al,E_\ga] = 0$.}
\begin{align*}
T_1 + i T_2 &= e_{12}, &
T_4 + i T_5 &= e_{13}, &
T_6 + i T_7 &= e_{23},
\\
T_9 + i T_{10} &= e_{14}, &
T_{11} + i T_{12} &= e_{24}, &
T_{13} + i T_{14} &= e_{34}.
\end{align*}
We keep the normalization $\tr(T_a T_b) = \half\,\dl_{ab}$. The
structure constants $f_{abc}$ have the following nonzero squares, with
a hexadecimal labelling:
\begin{gather}
(f_{123})^2 = 1; \quad
(f_{458})^2 = (f_{678})^2 = \tfrac{3}{4};
\nn \\
(f_{89A})^2 = (f_{8BC})^2 = \tfrac{1}{12}; \quad
(f_{8DE})^2 = \third;
\nn \\
(f_{147})^2 = (f_{156})^2 = (f_{19C})^2 = (f_{1AB})^2
= (f_{246})^2 = \quarter,
\nn \\
(f_{257})^2 = (f_{29B})^2 = (f_{2AC})^2 = (f_{345})^2 = (f_{367})^2
= \quarter,
\nn \\
(f_{39A})^2 = (f_{3BC})^2 = (f_{49E})^2 = (f_{4AD})^2 = (f_{59D})^2
= \quarter,
\nn \\
(f_{5AE})^2 = (f_{6BE})^2 = (f_{6CD})^2 = (f_{7BD})^2 = (f_{7CE})^2
= \quarter;
\nn \\
(f_{9AF})^2 = (f_{BCF})^2 = (f_{DEF})^2 = \tfrac{2}{3}.
\label{eq:su4-struct} % (34)
\end{gather}
Naturally, the structure constants for the Lie subalgebra $SU(3)$ are
a subset of those for $SU(4)$. Thus objections to putative models for
$SU(3)$ carry over to the $SU(4)$ case. Nevertheless, the allowable
pattern for $SU(3)$ given by \eqref{eq:dura-lex} does have an analogue
for $SU(4)$. Let us assume that the bosons labelled by the $SU(3)$
subalgebra are massless, and that the new ones are massive:
$$
m_1 =\cdots= m_8 = 0;  \qquad  m_9, \dots, m_{15} \neq 0.
$$
The relation \eqref{eq:the-thin-edge} together with 
\eqref{eq:su4-struct} gives at once
$$
m_9 = m_{10} = m_{11} = m_{12} = m_{13} = m_{14} =: m,
$$
but remains silent about $m_{15}$. Now we check this for consistency
with \eqref{eq:minas-gerais}. For $a \leq 8$, $b \geq 9$, this
relation always reduces to $0 = ((2m^2)^2 - 4m^4)/m^2$. Taking
$a \neq b$ in the range $\{9,\dots,14\}$ we typically obtain
\begin{equation}
4 m^2 (f^5_{aa})^2 = 4 m^2 \sum_{d\leq 8} (f_{abd})^2
+ (f_{abF})^2 (4 m^2 - 3 m_{15}^2).
\label{eq:res-ipsa-loquitur} % (35)
\end{equation}
In most cases, this reduces to $(f^5_{aa})^2 = \quarter$. When
$(a,b) = (9,10)$ or $(11,12)$ or $(13,14)$, we then get
$$
m^2 = \third(4 m^2) + \tfrac{2}{3} (4 m^2 - 3 m_{15}^2),
$$
that is, $3 m^2 = 2 m_{15}^2$. When $a = 15$ and $9 \leq b \leq 14$,
the constraint \eqref{eq:minas-gerais} becomes
$4 m^2 (f^5_{FF})^2 = (2 m_{15}^4 / 3 m^2) = m_{15}^2$, consistent 
with $f^5_{FF}/f^5_{bb} = m_{15}/m$, as expected. To sum up, this
mass pattern is compatible with the CGI mass conditions
\eqref{eq:madre-del-cordero} and~\eqref{eq:timo-del-siglo-xx},
provided that
$$
m_{15} = \sqrt{3/2}\, m.
$$
No other pattern for $SU(4)$ with one physical scalar seems to solve
\eqref{eq:madre-del-cordero} and~\eqref{eq:timo-del-siglo-xx},
although an exhaustive search has not been performed.

\smallskip

Going to the general $SU(n)$ case, one can show likewise that for a
theory with $n^2 - 1$ vector bosons and one physical scalar:
\begin{gather*}
m_1 = m_2 =\cdots= m_{n^2-2n} = 0;
\\
m_{(n-1)^2} =\cdots= m_{n^2-2} =: m \neq 0;
\\
\quad m_{n^2-1} = \sqrt{\frac{2(n-1)}{n}}\, m.
\end{gather*}
The ``odd man out'' corresponds to the last Cartan matrix
$$
T_R = C_n = \frac{1}{\sqrt{2n(n - 1)}}\, \diag(1, \dots, 1, -(n-1)),
$$
while the previous ones become massless. With the labels
$D = n^2 - 2n$, $P = n^2 - 3$, $Q = n^2 - 2$, $R = n^2 - 1$, then for
$(n - 1)^2 \leq a \leq n^2 - 3$ and $b = a + 1$, one finds that%
\footnote{To compute $f_{PQD}$ and $f_{PQR}$, just evaluate the 
commutators $[C_{n-1}, e_{n-1,n}]$ and $[C_n, e_{n-1,n}]$.}
\begin{align*}
\sum_{d\leq D} (f_{abd})^2
&= (f_{PQD})^2 = \frac{n - 2}{2n - 2}\,,
\\[\jot]
(f_{abR})^2 &= (f_{PQR})^2 = \frac{n}{2n - 2}\,.
\end{align*}
Thus, the analogue of \eqref{eq:res-ipsa-loquitur} for the $SU(n)$
case is
$$
m^2 = \frac{2(n - 2)\,m^2}{n - 1}
+ \frac{n}{2(n - 1)}\,(4 m^2 - 3 m_R^2),
$$
yielding
\begin{equation}
m_R^2 = \frac{2n - 2}{n}\, m^2.
\label{eq:lex-talionis} % (36)
\end{equation}
It seems clear that the masses of the gauge particles organize
themselves in $SU(n - 1)$ multiplets, concretely the fundamental one
and a singlet.

Thus the translation of our allowed models into the SSB phraseology
follows a well-trodden path: in general, a vector representation for
$SU(n)$ contains $2n$ real fields, of which $2n - 1$ are ``eaten'' to
provide the longitudinal components for $2n - 1$ ``initially
massless'' gauge fields, leaving $n^2 - 1 - (2n - 1) = (n - 1)^2 - 1$
``photons'' (corresponding to an $SU(n - 1)$ as yet ``unbroken''
symmetry), and the remaining one is the physical higgs field. In such
a framework formula~\eqref{eq:lex-talionis} is well known
\cite[Sect.~84]{Srednicki}. Things work out similarly for vector
representations of $O(n)$; the $O(3)\simeq SU(2)$ case we have seen
already in Sect.~\ref{ssc:SU2}.

In conclusion, the CGI mass conditions \eqref{eq:madre-del-cordero}
and~\eqref{eq:timo-del-siglo-xx} efficiently identify SSB-type models
in the vector representation. The reader will have no difficulty in
writing the $S$-representations and checking the commutation
relations.%
\footnote{Group theory dictates that the sum of the squared $f^5_{aa}$
be equal to the Casimir for the corresponding representations,
respectively~2 and~$\frac{3}{4}$ for the $SU(2)$ models just above.
For every $SU(n)$ model of this type $(f^5)^2 = \quarter$ holds.}

% \S 5.4
\subsection{Causal gauge invariance for $SU(3)$ with fields in the adjoint}
\label{ssc:suerte-o-verdad}

We finally turn to more involved models with the scalar fields in the
adjoint representation, to exemplify minimal coupling in the CGI
framework for such models and to deal with the alleged clash between
CGI and SSB in~\cite{FortunaJuvet}.

Concerning GUT models, (most simply) \textit{two} irredu\-cible
representations of pre-higgs particles are needed for SSB to yield
something recognizably akin to the SM; to wit, in~\cite{GGstrikeagain}
the adjoint representation $\mathbf{24}$ and the complex fundamental
representation $\mathbf{5}$ of $SU(5)$. Thus the question is whether a
model with 24 vector bosons, of which~12 have identical nonzero mass
and 12 are massless, passes muster in~CGI, allowing for 12 higgs-like
fields. Naturally, one should seek to answer the similar question for
simpler models first. For $SU(2)$, the model with 3 vector bosons, two
with identical nonzero mass and one massless, together with one
higgs-like field, has been shown in Sect.~\ref{ssc:SU2} to pass muster
in CGI. For $SU(3)$ there would be in the adjoint representation 8
vector bosons, 4~of which have identical nonzero mass and 4 are
massless, with 4 higgs-like fields. We take up this case as the
simplest proxy for our problem, embarking on this from the opposite
end to that of Sect.~\ref{ssc:Abel}: first we recall SSB for this
model; then in consonance with \cite{Michael05} we verify BRST
invariance for the resulting classical Lagrangian; this proves CGI at
tree level. Finally, we check the representation property
\eqref{eq:addio} of the $S$-matrices, and substantiate our claim in
Sect.~\ref{ssc:a-la-vaca} that minimal coupling follows from CGI. We
deem the exercise important and proceed in fastidious detail.

% \S 5.4.1
\subsubsection{BRST invariance of the classical Lagrangian}
\label{sss:unto-the-breach}

An invariant potential for that representation is
\begin{equation}
V(\Phi) = -\mu^2 \tr\Phi^2 + \la (\tr\Phi^2)^2
\label{eq:end-of-the-beginning} % (37)
\end{equation}
with $\Phi$ a traceless hermitian $3 \x 3$ matrix variable and
$\la > 0$. (We are not striving for maximum generality here, so we
have suppressed a term in $\tr\Phi^3$. The usual $\tr\Phi^4$ term is
missing since in this somewhat degenerate case
$\tr\Phi^4 = \half(\tr\Phi^2)^2$ by the Cayley--Hamilton formula.)
The pattern of symmetry breaking is $S(U(2) \x U(1))$; a minimum of
the potential $V(\Phi)$ is of the form
$$
\val{\Phi} = v \diag\bigl( 1/2\sqrt3, 1/2\sqrt3, -1/\sqrt3 \bigr)
= v\la_8/2,
$$
where $v$ is to be determined such that $\widetilde V(v) :=
V(v\la_8/2) = -\mu^2 v^2/2 + \la v^4/4$ be minimal, see
\cite{Rafferty} for instance. This gives $v^2_{\min} =
\frac{\mu^2}{\la}$; $\widetilde V(v_{\min}) = - \frac{\mu^4}{4\la}$.
From now on, we just write $v$ for~$v_{\min}$. Also write $A^\mu
\equiv A^\mu_a T_a$, $\Phi \equiv \phi_a T_a$, $u \equiv u_a T_a$,
$\ut \equiv \ut_a T_a$, using the Gell-Mann basis. One can easily
check that a shifted field $\vf$ is required only for the $\phi_8$
component: $\phi_8 = v + \vf$.

The covariant derivative in the adjoint representation is of the form
$D^\mu = \del^\mu - i\ka\,[A^\mu,\,\cdot\,]$; in components,
\begin{align*}
D^\mu_{ab} &= \dl_{ab}\,\del^\mu - \ka\,f_{abc}\,A^\mu_c,
\word{and thus}
\\[\jot]
D^\mu \Phi &= D^\mu(\phi_e T_e) 
= \del^\mu \phi_e\,T_e - i\ka\,[A^\mu_a T_a, \phi_b T_b]
\\[\jot]
&= \del^\mu \phi_e\,T_e + \ka f_{abc} A^\mu_a \phi_b T_c
= (\del^\mu \phi_b + \ka f_{abc} A^\mu_c \phi_a) T_b
\\[\jot]
&= (\dl_{ab}\, \del^\mu \phi_b - \ka f_{abc} A^\mu_c \phi_b) T_a
= (D^\mu \Phi)^\7.
\end{align*} 
 
The Lagrangian for bosonic scalar fields reads
$$ 
L_\Phi = \tr(D\Phi \. D\Phi) - V(\Phi), 
$$ 
where $V$ is given by \eqref{eq:end-of-the-beginning} with
$\mu^2 \sim \ka^0$ and $\la \sim \ka^2$. To determine the mass
spectrum of the gauge fields one collects the mass terms in
$\tr(D\Phi\.D\Phi)$, with $\phi_8$ replaced by $v + \vf$, with the
result that
\begin{align}
& \ka^2 (A_b\. A_d)\, f_{8ab} f_{8cd}\,v^2 \tr(T_a\,T_c) 
\nn \\
&= \frac{3v^2\ka^2}{8} 
(A_4\. A_4 + A_5\. A_5 + A_6\. A_6 + A_7\. A_7).
\label{eq:son-tuyos} % (38)
\end{align}
Hence $m_1^2 = m_2^2 = m_3^2 = m_8^2 = 0$, and 
$$
m^2 := m_4^2 = m_5^2 = m_6^2 = m_7^2 = 3\mu^2\ka^2/4\la = 3v^2\ka^2/4.
$$
The potential $V(\Phi)$ contains a mass term only for the field~$\vf$,
namely one can show that $V(\Phi) = V_0 + \mu^2\vf^2 + {}$(terms
trilinear and quadrilinear in the fields), where
$V_0 := \widetilde V(v_{\min})$. Hence $\vf$ is the ``polar'' higgs
field, in the direction of symmetry breakdown. The other three higgs
fields are massless, they are pseudo-Goldstone bosons in the precise
sense of~\cite{pseudoWeinberg}. Next we collect all terms $\sim \ka^0$
in $\tr(D\Phi\. D\Phi)$. Since there appears a term
\begin{align*}
& 2\tr\bigl[ (\del_\mu \phi_a T_a)(\ka f_{8bc}\,A^\mu_c\,v T_b) \bigr]
\\
&\quad = \frac{2m}{\sqrt{3}} \del_\mu \phi_b f_{8bc}\,A^\mu_c
= -\frac{2m}{\sqrt{3}} \del_\mu \phi_b f_{ab8}\,A^\mu_a,
\end{align*} 
we introduce the notation
$$
B_a = \frac{2}{\sqrt{3}}\,f_{ab8}\,\phi_b,
$$
that is, $B_4 = \phi_5$, $B_5 = -\phi_4$, $B_6 = \phi_7$,
$B_7 = -\phi_6$, and $B_a = 0$ for $a = 1,2,3,8$; later we shall see
that the $B_a$'s are the St\"uckelberg fields belonging to the massive
vector bosons $A_a$ (for $a = 4,5,6,7$). Therefore the remaining
bosonic scalar fields $\vf_1 \equiv \phi_1$, $\vf_2 \equiv \phi_2$,
$\vf_3 \equiv \phi_3$ are the massless physical higgs fields. With
that one obtains
\begin{align*}
\tr(D\Phi \. D\Phi)
&= \frac{1}{2} \sum_{a=1,2,3} \del\vf_a \. \del\vf_a
+ \frac{1}{2} \sum_{a=4,5,6,7} \del B_a \. \del B_a
\nn \\
&\quad + \frac{1}{2} \del\vf \. \del\vf
+ \frac{m^2}{2} \sum_{a=4,5,6,7} (A_a \. A_a)
\nn \\
&\quad - m \sum_{a=4,5,6,7} A_a \. \del B_a + O(\ka).
\end{align*}

We now assemble the last pieces of the Lagrangian. For the $mA\.\del B$
terms to add up to a divergence, choose for the gauge-fixing term in
the Feynman gauge
\begin{align*} 
L^\gf &= L_0^\gf = -\frac{1}{2} \sum_a (\del\. A_a + m\,B_a)^2 
\nn \\ 
&= -\frac{1}{2} \sum_{a=1}^8 (\del\. A_a)^2 
- \frac{m^2}{2} \sum_{a=4,5,6,7} B_a^2 
\nn \\ 
&\quad - m \sum_{a=4,5,6,7} (\del\. A_a) B_a. 
\end{align*} 
Due to $F^{\mu\nu} \equiv \del^\mu A^\nu - \del^\nu A^\mu
- i\ka\,[A^\mu, A^\nu]$, the Yang--Mills Lagrangian 
$$
L^\YM = -\frac{1}{2} \tr F^{\mu\nu} F_{\mu\nu}
= -\frac{1}{4} F^{\mu\nu}_a \,F_{a\,\mu\nu}
$$
is of the form $L^\YM = L_0^\YM + \ka\, L_1^\YM + \ka^2\, L_2^\YM$.

\medskip

Next we introduce the BRST transformation and verify classical BRST
invariance of the total Lagrangian.%
\footnote{The reader might ask: why bother? Are not all models
generated by SSB automatically BRST invariant? Surely yes. But there
are few discussions of this in the literature: typically textbooks go
at great length into the proof that Yang--Mills theories are BRST
invariant, and then resolutely tiptoe around the same question for
``hidden local symmetry''. Reference~\cite{VacuumCleaner} furnishes an
amusing example.}

For $A^\mu$ and $\Phi$ the $s$-operator is given by the infinitesimal
gauge transformations
$$
s A^\mu = D^\mu u = \del^\mu u - i\ka\,[A^\mu, u],  \quad
s \Phi = i\ka\,[u,\Phi].
$$
For example, the latter gives
\begin{align*}
s \vf_ 1 &= -\ka(u_2 \vf_3 - u_3 \vf_2)
\\
&\quad - \frac{\ka}{2} (u_4 B_6 + u_7 B_5 + u_6 B_4 + u_5 B_7);
\end{align*}
in general, $s \vf_a = O(\ka)$ for $a = 1,2,3$; whereas
\begin{align*}
s B_4 
&= m\,u_4 + \frac{\ka}{2} (u_6 \vf_1 + u_1 B_7 - u_7 \vf_2 + u_2 B_6
\\
&\quad + u_3 B_5 + u_4 \vf_3 + \sqrt{3}(u_4 \vf + u_8 B_5));
\end{align*}
in general, $s B_a = m\,u_a + \Oh(\ka)$ for $a = 4,5,6,7$; and
$$
s \vf = - \frac{\ka\,\sqrt{3}}{2} \sum_{a=4,5,6,7} u_aB_a.
$$

For the ghost fields $u$, we get $su = i\,\ka/2 \,[u,u]$, as in the
massless case, and for the antighosts $\ut$,
$$
s \ut = -(\del\. A + m\,B).
$$
With that the gauge fixing term can be written as
$L_\gf = -\half \sum_a (s\ut_a)^2$, as usual in the Feynman gauge.

For the nilpotence of $s$, we find that $s^2 A = 0$ and $s^2 u = 0$
are similar to the well-known massless case. In the Lie superalgebra
generated by the $A = A_a^\mu T_a$ and the $u = u_b T_b$, we can write
\begin{align*}
& s^2 A = s(Du) = \del(s u) - i\ka[s A, u] - i\ka[A, s u]
\\
&\quad = \frac{i\,\ka}{2} \bigl( [u, \del u] - [\del u ,u] \bigr) 
- \frac{\ka^2}{2} \bigl( 2\,[[A,u], u] - [A, [u,u]] \bigr).
\end{align*}
Since the bracket is symmetric between expressions of ghost number~1,
the first two terms cancel; and the second two terms also cancel by
the same symmetry and the Jacobi identity, whereby $[A,\cdot]$ is a
derivation:
$$
[A, [u,u]] = [[A,u], u] + [u, [A,u]] = 2 [[A,u], u].
$$

The same identity takes care of $s^2 \Phi$:
\begin{align*}
s^2 \Phi &= i\ka\,\bigl( [s u, \Phi] - [u, s \Phi] \bigr)
\\
&= - \frac{\ka^2}{2} \bigl( [[u,u], \Phi] - 2[u, [u,\Phi]] \bigr) = 0.
\end{align*}
Vanishing of $s^2\ut$ is discussed further down.

The BRST transformation of $A^\mu$ and $\Phi$ has the form of an
infinitesimal gauge transformation of the unshifted field. It follows
that $sL^\YM = 0$ and $sL_\Phi = 0$. We perform the explicit
verification for $L_\Phi$ by using:
\begin{align*}
s(D\Phi) &= \del(s \Phi) - i\ka \bigl([s A, \Phi] + [A, s \Phi]\bigr)
\\
&= i\ka\,\del([u, \Phi]) - i\ka[\del u, \Phi]
\\
&\quad + \ka^2 \bigl( -[[A, u], \Phi] + [A, [u, \Phi]] \bigr)
\\
&= i\ka [u, \del\Phi] + \ka^2 [u, [A, \Phi]] = i\ka [u, D\Phi].
\end{align*}
Since $D\Phi = \del\Phi - i\ka [A, \Phi]$ has ghost number zero, it 
follows that
\begin{align*}
s\,\tr(D\Phi \. D\Phi)
&= i\ka \tr\bigl( [u, D\Phi] \. D\Phi + D\Phi \. [u, D\Phi] \bigr)
\\
&= i\ka \tr[u, D\Phi \. D\Phi] = 0.
\end{align*}
Next we introduce the ghost Lagrangian, chosen in a such a way that
$s L^\gf + s L^\gh$ is a divergence: one sets
$$
L^\gh = \sum_{a=1}^8 \del_\mu \ut_a\,s A^\mu_a
- m \! \sum_{a=4,5,6,7} \ut_a\,s B_a = L_0^\gh + \ka L_1^\gh,
$$
which yields indeed $s(L^\gf + L^\gh) = \del\. I$, with
\begin{equation}
I := -(\del\. A_a + m\,B_a) \,D_{ab} u_b = I_0 + \ka\, I_1.
\label{eq:I-term} % (39)
\end{equation}

Summing up, the total Lagrangian
\begin{equation}
L_\total = L^\YM + L_\Phi + L^\gf + L^\gh
= - V_0 + L_0 + \ka\ L_1 + \ka^2 L_2
\label{eq:Lagr-total} % (40)
\end{equation}
is classically BRST invariant since
$$
s L_\total = \del \. I.
$$
All terms of $L_2$ come from $L^\YM + L_\Phi$. Notice that
\begin{align} 
L_0 &= \sum_{a=1}^8 \biggl(-\frac{1}{4}(A_a^{\mu,\nu} - A_a^{\nu,\mu})
(A_{a\,\mu,\nu} - A_{a\,\nu,\mu}) 
\nn \\
&\quad - \frac{1}{2}\, \del A_a \. \del A_a \biggr)
+ \frac{m^2}{2} \! \sum_{a=4,5,6,7} A_a \. A_a
\nn \\ 
&\quad + \sum_{a=1}^8 \del \ut_a \. \del u_a
- m^2 \! \sum_{a=4,5,6,7} \ut_a u_a
\nn \\
&\quad - m \! \sum_{a=4,5,6,7} \del\. (A_a\,B_a)
+ \frac{1}{2} \biggl( \sum_{p=1,2,3} \del\vf_p \. \del\vf_p 
\nn \\ 
&\quad 
+ \! \sum_{a=4,5,6,7} (\del B_a \. \del B_a - m^2\,B_a^2)
+ \del\vf \. \del\vf - 2\,\mu^2 \vf^2 \biggr)
\label{eq:Lzero-massive} % (41)
\end{align} 
contains the divergence term $-m\,\del\.(AB)$, which is irrelevant for
the field equations, but contributes to the BRST current $j_{(0)}$ of
the free theory (see below).

We may go back now to nilpotence of~$s$. The vanishing of $s^2 \ut_a$
takes place on-shell: with $S_\total = \int d^4 x\, L_\total(x)$, the
relation
$$
s^2\ut_a = \frac{\dl\,S_\total}{\dl\,\ut_a} = 0
$$
follows from the Euler--Lagrange equations.

\smallskip

The BRST current of the free theory can be computed by the formula:
\begin{equation}
j_{(0)}^\mu = -\biggl( \frac{\del L_0}{\del(\del_\mu\vf_i)}\, s_0\vf_i
+ I_0^\mu \biggr),
\label{eq:free-current} % (42)
\end{equation}
where there is a summation over $\vf_i = A^\mu_a$, $u_a$, $\ut_a$,
$B_a$, $\vf_p$, $\vf$. For the present model it comes out that
\begin{align*}
j_{(0)}^\mu &= (\del\. A_a + m_a\,B_a) \del^\mu u_a
- u_a\,\del^\mu(\del\. A_a + m_a\,B_a)
\\
&\quad - \del_\nu( (\del^\nu A^\mu_a - \del^\mu A^\nu_a) u_a)
+ u_a(\square + m_a^2) A_a^\mu,
\end{align*}
by inserting the explicit expressions in~\eqref{eq:I-term} for $I_0$
and \eqref{eq:Lzero-massive} for $L_0$ into \eqref{eq:free-current}.
As explained in~\cite{Michael05}, the last two terms do not contribute
to the nilpotent charge $Q_\mathrm{in}$ ---which is defined only for
on-shell fields. Hence, the superderivation
$[Q_\mathrm{in},\cdot]_\mp$ is defined as in~\cite{Scharf}, and
the $B_a$ are indeed the St\"uckelberg fields of this reference.

Since the second CGI criterion~\cite{Michael05} is fulfilled, the
model is causal gauge invariant at the tree level to all orders.

% \S 5.4.2
\subsubsection{Minimal coupling from CGI with fields in the adjoint}
\label{sss:pro-domo-sua}

\renewcommand{\arraycolsep}{3pt} %% narrower matrix columns here

Let us now take stock of the CGI relation~\eqref{eq:addio} for the
adjoint $SU(3)$ model. In order to avoid misunderstandings, we remind
the reader that each vector boson is entitled to its matrix, but there
are no rows or columns corresponding to the massless ones. The
labelling of the entries is thus given in the order: $4,5,6,7$ (for
the MVB) and $1,2,3,8$ (for the ``higgses''). That is to say, in the
notation of~\eqref{eq:eta}, we use the multiplet
\begin{equation}
\eta = \begin{pmatrix} B_4 \\ B_5 \\ B_6 \\ B_7 \\
\phi_1 \\ \phi_2 \\ \phi_3 \\ v + \vf \end{pmatrix}
\word{versus}
\begin{pmatrix} \phi_1 \\ \phi_2 \\ \phi_3 \\ \phi_4 \\
\phi_5 \\ \phi_6 \\ \phi_7 \\ \phi_8 \end{pmatrix} 
= \begin{pmatrix} \vf_1 \\ \vf_2 \\ \vf_3 \\
-B_5 \\ B_4 \\ -B_7 \\ B_6 \\ v + \vf \end{pmatrix}
\label{eq:vamos-Rafa} % (43)
\end{equation}
of the previous subsection. To write down the matrices $S^a$, we read
off $f^3_{***}$, $f^5_{***}$, $f^6_{***}$ from 
$\ka\,f_{abc}\,(A_c\.\del\phi_b)\,\phi_a$ in $\tr(D\Phi\.D\Phi)$; note
that \eqref{eq:Lagr-total} contains no further contributions to these
cubic coupling coefficients. With that, we obtain
$$
S^1 = \begin{pmatrix}
&& 0 & -\half \\ && \half & 0 \\ 0 & -\half \\ \half & 0 \\
&&&& 0 \\ &&&&& 0 & -1 \\ &&&&& 1 & 0 \\ &&&&&&& 0 \end{pmatrix}.
$$
Here we note at once the new relations: 
$-2f^6_{123} = 2f^6_{132} = -1$ (so $f^6$ is not zero). The
contributions in the upper left corner come from $f^3_{147}$ and such,
that do not vanish. Similarly:
\begin{align*}
S^2 &= \begin{pmatrix}
&& -\half & 0 \\ && 0 & -\half \\ \half & 0 \\ 0 & \half \\
&&&&&& 1 \\ &&&&& 0 \\ &&&& -1 \\ &&&&&&& 0 \end{pmatrix};
\\[\jot]
S^3 &= \begin{pmatrix}
0 & -\half \\ \half & 0 \\ && 0 & \half \\ && -\half & 0 \\
&&&& 0 & -1 \\ &&&& 1 & 0 \\ &&&&&& 0 \\ &&&&&&& 0 \end{pmatrix};
\end{align*}
and the commutation relations of the $SU(2)$ subgroup are clearly
fulfilled.

The reader will easily write down the other basis matrices of the
$S$-representation and check the group property. The main remark is
that, besides the nonvanishing of $f^6_{1**}$, $f^6_{2**}$,
$f^6_{3**}$, there are nondiagonal values of $f^5_{**p}$, for example
$f^5_{472} = -f^5_{461} = \half$. Matters work in the same way as in
this example for any $SU(n)$ model in the adjoint representation for
higher~$n$. Look no farther for the solution to the conundrum raised
by \cite{FortunaJuvet}: the model coming by SSB of the representation
$\mathbf{24}$, responsible for ``superstrong breaking'' in the
original GUT by Georgi and Glashow, is incompatible with the
obstructions given by Scharf~\cite{Scharf}; but CGI does not exclude
it.

We shall now verify that the scalar-gauge Lagrangian resulting from
the first CGI method can be expressed as the minimal coupling
$\half (D\eta)^t\. D\eta = \half \sum_b (D\eta)_b\. (D\eta)_b$. The
former can be expressed by the term
\begin{align*}
\tr(D\Phi \. D\Phi) &= \frac{1}{2} \sum_b (D\phi)_b \. (D\phi)_b,
\\
(D\phi)_b &:= \del\phi_b - \ka\,f_{bac}\,A_c\,\phi_a
\end{align*}
obtained by SSB, since construction of the $SU(3)$-adjoint model by
SSB agrees with what one obtains by CGI (as shown in
Sect.~\ref{ssc:suerte-o-verdad}). Then
$\half(D\eta)^t\. D\eta = \tr(D\Phi\. D\Phi)$ follows from noting that
the sets of covariant derivatives agree. For example, by direct
calculation,
\begin{align*}
(D\phi)_1 &= \del\vf_1 + \ka(A_2\vf_3 - A_3\vf_2)
\\
&+ \frac{\ka}{2} (A_4 B_6 + A_6 B_4 + A_5 B_7 + A_7 B_5) = (D\eta)_5,
\\
(D\phi)_4 &= -\del B_5 - \frac{\ka}{2}
(A_1 B_6 - A_2 B_7 + A_3 B_4 - A_5 \vf_3 
\\
- A_6 \vf_2 
&- A_7 \vf_1) - \frac{\ka\sqrt{3}}{2} (-A_5(\vf + v) + A_8 B_4)
= -(D\eta)_2,
\\
(D\phi)_8 &= \del\vf + \frac{\ka\sqrt{3}}{2}
(A_4 B_4 + A_5 B_5 + A_6 B_6 + A_7 B_7) 
\\
&= (D\eta)_8.
\end{align*}
Agreement of other components follows by permutation.

\smallskip 

It is instructive to write down term by term the equality
$\half (D\eta)^t\. D\eta ={}$ scalar gauge Lagrangian obtained by the
first CGI method, see~\eqref{eq:mini-couple}. The validity of the
resulting equations is not limited to the $SU(3)$-adjoint model.

The mass terms for vector gauge fields originate from
$(A_a\. A_b)\,v^t\,[S_a, S_b]_+ v$, where $v^t := (v_1,\dots,v_z)$.
Equating the corresponding coefficients, we find that
$$
-\frac{\ka^2}{4}\,v^t\,
\bigl( -{}^tG^aG^b - {}^tG^b G^a + [H^a,\,H^b]_+ \bigr)\,v 
= \delta_{ab}\, m_a^2.
$$
For the $SU(3)$-adjoint model, with the scalar multiplet as
in~\eqref{eq:vamos-Rafa}, on inserting the above $S$-matrices, this
formula gives indeed $\frac{3}{8} v^2 \ka^2 \sum_{k=4}^7 A_k\. A_k =
\half m^2 \sum_{k=4}^7 A_k\. A_k$, like in~\eqref{eq:son-tuyos}. Note
that although $f^6 \neq 0$, it does not contribute.

By the definition of the $S$-matrices, the couplings
$L_1^3 + L_1^5 + L_1^6$ must be contained in
$(A_a\.\del\eta^t\,)S^a\eta - \eta^t S^a(A_a\.\del\eta)$. Indeed, we
obtain
\begin{align*}
& (A_a \.\del\eta^t\,) S^a \eta - \eta^t S^a (A_a \.\del\eta)
\\
&\quad = L_1^3 + L_1^5 + L_1^6 - \ka\,f^5_{abp}\,(A_a\.\del B_b)\,v_p.
\end{align*}
The additional $(A\.\del B)$ term gives, as expected,
\begin{align*}
-\ka f^5_{abp} (A_a\.\del B_b) v_p &= -\ka v f^5_{ab8} (A_a\.\del B_b)
\\
= -\frac{\sqrt3\,\ka v}{2}\! \sum_{a:\,m_a\neq 0} (A_a\.\del B_a)
&= -m\!\sum_{a:\,m_a\neq 0} (A_a\.\del B_a).
\end{align*}

Moreover, in $(A_a\. A_b)\eta^t\,[S_a,\,S_b]_+ \eta$ there are
trilinear terms corresponding to $L_1^2 + L_1^7$. Equating the
pertinent coefficients, we get
\begin{align*}
f^2_{ab\star}
&= -\frac{\ka^2}{2}\,(F^a G^b + F^b G^a + G^a H^b + G^b H^a)v;
\\
f^7_{ab\star}
&= \frac{\ka^2}{2}\,({}^tG^a\,G^b +{}^tG^b\,G^a - H^a H^b - H^b H^a)v.
\end{align*}

The pattern exemplified above is typical for $SU(n)$ models with the
scalar fields in the adjoint representation. An even simpler example
is provided by the three-boson case with two identical masses and one
``photon'' of Sect.~\ref{ssc:SU2}. Proceeding as above, the reader
should have no difficulty in verifying that $\half(D\eta)^t\. D\eta$
equals the SSB-type expression $\tr(D\Phi\. D\Phi)$.

% \S 6
\section{Conclusion}
\label{sec:summa}

The Higgs sector of the SM, in the perspective of basic structures of
gauge theories, plays a somewhat ambiguous and enigmatic role. The
massless and massive gauge bosons which are the carriers of the
fundamental forces belong to what might be termed \textit{radiation},
in analogy to electrodynamics. Now, by itself, the gauge boson sector
of gauge theories of interest for physics defines a nontrivial theory.
Quarks and leptons, on the other hand, belong to the category
\textit{matter} which cannot ``live on its own'' without the gauge
sector. Indeed, a theory of quarks and leptons only is a theory of
free particles and, being untestable in experiment, is uninteresting.
The Higgs sector's place in this classification is perhaps not as
obvious as it may appear at first sight. Extensions of the Standard
Model within noncommutative geometry
\cite{CoLo,CoEFSch,Zappafrank,ConnesMix} view scalar fields as an
integral part of the connection, i.e. classify them in the sector of
gauge bosons and hence place them in the category ``radiation''. Its
alleged phenomenological role of providing masses for the fermions and
some bosons of the model, and its likely kinship with dark
matter~\cite{DarkSide}, in turn, might suggest that it rather
represents another form of ``matter'' beyond the ordinary one made out
of quarks and leptons.

Be that as it may, the traditional description of the Higgs sector by
means of the ``hidden symmetry'' concept, however attractive it may
seem from the standpoint of group theory, is still a purely
\textit{classical} one. Classical and semi-classical mechanisms have
their uses in quantum field theory: no one will dispute that anomalies
are a quantum phenomenon although they can be described in purely
classical terms \cite{Heil-Eins,Heil-Zwei}. For conceptual clarity,
nevertheless, one should cling to root quantum explanations.

One may reckon, furthermore, that on the subject of this paper the
panorama has been obscured by much theoretical prejudice. The Higgs
mechanism is burdened with giving masses to \textit{all} matter and
force fields; a heavy load to carry indeed. Explicit mass terms for
the vector bosons of electroweak theory are said to be forbidden by
gauge invariance. It ain't so: these mass terms can be accommodated in
gauge theory by regarding the ``swallowed'' Higgs ghosts of lore as
St\"uckelberg fields. Also it is said that chirality of the fermions
and gauge invariance in weak interactions requires the Higgs mechanism
to generate masses by Yukawa couplings. It ain't so: one can use Dirac
masses for the fermions and derive chirality of couplings from causal
gauge invariance~\cite{PGI-EW-I,PGI-EW-II}.

In conclusion, starting from the BRST description for MVB as
fundamental objects
\cite{PGI-EW-I,CabezondelaSal,Felicitas,Scharf,LuisDixit}, we have
perturbatively performed a second reality check of the Higgs mechanism
in the spirit of causal gauge invariance, with the outcome that,
reversing the \textit{dictum} by Yang, \textit{interaction dictates
symmetry} ---fixing the models up to minute details.%
\footnote{A role for MVB as sources of symmetry, with very different
intent, is found in~\cite{ChacraRanaLoco}.}
That vindicates the conclusions of the historically first reality
check \cite{CLT74,LT75} as~well. Beyond re\-establishing the manifold
aspects of renormalizable gauge theories, the analysis in the
path-breaking book~\cite{Scharf} has been completed with the causal
derivation of minimal coupling. This allows now for a reliable list of
renormalizable couplings in BRST~invariant models. The contention that
there might be contradiction between causal gauge inva\-riance and
some GUT models~\cite{FortunaJuvet} has been laid to rest.

\appendix

\section*{Appendices} %% (unlabelled)
\addcontentsline{toc}{section}{\protect\numberline{}Appendices}

% \S A
\section{On the Standard Model in CGI}
\label{app:SM}

Postulating four gauge bosons, one of which is massless, and one
physical scalar, one is unerringly led by CGI~\cite{PGI-EW-I} to 
$U(2)$ symmetry%
\footnote{As remarked early on in~\cite{Oldebook}, the true group of
the electroweak interaction is~$U(2)$, not $SU(2)\x U(1)$.}
and the ``phenomenological'' boson sector of the~SM. The only
alternatives allowed by CGI are limits of the SM in which one, three
or all of the vector particles decouple. In standard presentations the
$U(2)$ symmetry is said to be ``broken'', among other reasons, because
there is only one conserved quantity, electric charge, instead of
four. But from our viewpoint symmetry is broken at the level of the
free Lagrangian, due to different masses (the residual equality of two
masses reflects conservation of electric charge). This is to say that
there is a natural basis of the Lie algebra linked to the pattern of
masses. The role of the constraint \eqref{eq:timo-del-siglo-xx} is
precisely to pick out this basis. Little support comes from this
quarter for the idea that the SM as it stands is ``imperfectly
unified''.

\smallskip

Now, the CGI conditions can likewise be applied to the \textit{fermion
sector}. As hinted at earlier, incompatibility of Dirac masses for
fermions with gauge symmetry is just another popular misconception.
The basic fermion-vector-boson vertices between carriers and matter in
a gauge theory are written $\ka(b^a\,\bar\psi \Aslash_a \psi +
{b'}^a\,\bar\psi \Aslash_a \ga^5 \psi)$, \textit{\`a la} Bjorken and
Drell, with $\bar\psi$ the Dirac adjoint spinor and $b,b'$ appropriate
coefficients. Taking for the fermions the known ones ---see
\cite{PGI-EW-I,PGI-EW-II} and \cite[Sect.~4.7]{Scharf}--- first-order
gauge invariance already determines some couplings: in particular, the
photon has no axial vector couplings ``because'' there is no
St\"uckelberg field for it. At second order, contractions between the
corresponding fermionic $Q$-vertex and the bosonic $L_1$ and between
the bosonic $Q$-vertex and $L_1^F$ \textit{determine the matter
couplings} completely (contractions between $L_1^F$ and its $Q$-vertex
contribute nothing). It is beautiful to behold that couplings of the
physical scalar to fermions are proportional to their mass, and that
chirality of the interactions need not to be brought from the outside,
but is a consequence of~CGI. As usual, CGI at third order for tree
graphs fixes the higgs potential~\cite{PGI-EW-II}. Since the causal
version of the SM leads to the same phenomenological Lagrangian,
excepting only that the vacuum expectation value of the higgs field is
zero, there is no way within pure particle physics to tell it apart
from the ordinary version. However, all the above springs just from
the BRST treatment for free spin~one bosons and causal renormalization
theory. This stands our approach in good stead in the face of
breakdown of symmetry.

% \S B
\section{Derivation of the second main constraint}
\label{app:constraints}

The three constraints \eqref{eq:mater-et-magistra} are crucial in this
paper. All of them follow from~\eqref{eq:CGI-2}, that is, CGI for
second-order tree diagrams by using the technique explained in
Sect.~\ref{ssc:gore}. The third constraint is derived like the first,
except that the roles of the $B$- and $\vf$-fields are reversed; and
for the first one no substantial deviation from
reference~\cite{Scharf} is required. Thus we focus on the second
constraint.

Since only those terms of $P^\nu$ having a derivative $\del^\nu$ on a
field operator contribute, we list only such terms:
\begin{align*}
P^\nu &= \ka\bigl( f_{abc}(-A_{\mu a} u_b\,\del^\nu A^\mu_c
+ \half u_a u_b\,\del^\nu \ut_c)
\\
&\quad - 2\,f^3_{abc}\,u_a B_b\,\del^\nu B_c
\\
&\quad - f^5_{abp}\,u_a(B_b\,\del^\nu \vf_p - \del^\nu B_b\,\vf_p) 
\\
&\quad - 2\,f^6_{apq}\, u_a\vf_p\,\del^\nu\vf_q +\cdots \bigr).
\end{align*}
See (4.3.17) in \cite{Scharf} in this respect. The $k$-th term in this
expression will be called $P_k$ ($k = 1,\dots,6$) henceforth. We omit
the notation for normal ordering.

\smallskip

The second constraint is obtained from~\eqref{eq:CGI-2} by equating
the coefficients of $\dl(x - y)\,u_a u_b \ut_d \vf_p$.

A type~1 term $N_2 = C_{abdp}\,\dl\,B_a u_b \ut_d \vf_p$ would
contribute. However, as indicated in Sect.~\ref{ssc:a-la-vaca}, there
are no quartic terms containing ghost fields $u,\ut$. Also, there are
no type~2 or type~4 terms, since the contributions of these terms are
$\sim\!(\del\phi_1) \phi_2 \phi_3 \phi_4$ and not
$\sim\! \phi_1 \phi_2 \phi_3 \phi_4$.

The following type~3 terms do contribute. For the contraction of
$\del\ut$ in~$P_2$ with $u$ in~$L_1^8$, we must be careful with the
sign, because of the many Fermi operators. In the region $x^0 > y^0$
we obtain
\begin{align*}
& \T_2(P_2^\nu(x) \,L_1^8(y)) 
\\
&\quad = P_2^\nu(x) \,L_1^8(y)
\sim \wick:u_a u_b\,\del^\nu\ut_c(x): \,\wick:\ut_d u_{c'} \vf_p(y):
\\
&\quad = -i\,\dl_{cc'} \,\del^\nu_x \Delta_m^+(x - y)\,
\wick:u_a(x) u_b(x) \ut_d(y) \vf_p(y): +\cdots
\end{align*} 
and for $y^0 > x^0$,
\begin{align*}
& \T_2(P_2^\nu(x) \,L_1^8(y)) 
\\
&\quad = L_1^8(y) \,P_2^\nu(x)
\sim \wick:\ut_d u_{c'} \vf_p(y): \,\wick:u_a u_b\,\del^\nu\ut_c(x):
\\ 
&\quad = -i\,\dl_{cc'} \,\del^\nu_x \Delta_m^+(y - x)\,
\wick:u_a(x) u_b(x) \ut_d(y) \vf_p(y): +\cdots
\end{align*} 
by using Wick's theorem. Together these give a term
\begin{align*}
&\sim -i\,\del^\nu_x \bigl( \Delta_m^+(x - y) \theta(x^0 - y^0)
+ \Delta_m^+(y - x) \theta(y^0 - x^0) \bigr)
\\
&= -i \del^\nu \Delta_m^F(x - y).
\end{align*} 
On computing the divergence $\del_\nu^x$ and adding the term with $x$
and~$y$ exchanged, we find the contribution
$$ 
-i\, f_{abc} f^8_{dcp}\, u_a(x) u_b(x) \ut_d(x) \vf_p(x) \,\dl(x - y).
$$
Additional terms come from contracting $\del\vf$ in~$P_6$ with $\vf$
in~$L_1^8$ and $\del B$ in~$P_5$ with $B$ in~$L_1^4$. These
respectively read
\begin{align*}
-2i (f^6_{apv} f^8_{dbv} - f^6_{bpv} f^8_{dav})\,
u_a(x) u_b(x) \ut_d(x) \vf_p(x) \,\dl(x - y);
\\
-i(-f^5_{akp} f^4_{dbk} + f^5_{bkp} f^4_{dak})\,
u_a(x) u_b(x) \ut_d(x) \vf_p(x) \,\dl(x - y).
\end{align*}
On adding all terms and setting the resulting coefficient equal to
zero, the second constraint \eqref{eq:mater-et-magistra} follows.

% \S C
\section{Epistemological second thoughts}
\label{app:GuerraLira}

Among the motivations of this article was the realization of how
relatively poor a reputation SSB enjoys among knowledgeable
philosophers of science. In such quarters it is regarded as a
non-empirical device of little explanatory value. More precisely,
Higgs' argument is rightly seen as possessing tremendous heuristic
value in the \textit{context of discovery}, but less so in the context
of justification.
\begin{quote}
``As the semi-popular presentations put it, `particles get their
masses by eating the higgs field.' Readers of
\textit{Scientific\,American} can be satisfied\,with these just-so
stories. But philosophers of science should not be. For a genuine
property like mass cannot be gained by eating descriptive fluff, which
is just what gauge is. (They) should be asking\dots\ what is the
objective (i.e., gauge invariant) structure of the world corresponding
to the gauge theory presented in the Higgs mechanism?''
\end{quote}
This criticism by Earman is quoted in~\cite{NoClothesKing}, which
tries to explore the epistemological meaning of SSB. Consult as
well~\cite{HombreOreja}. A final remark is in order. When constructing
via CGI the Higgs potentials~$V$, a zero vacuum expectation value
emerges. Making this explicit is however noxious to the Higgs
mechanism interpretation. On which interpretation is preferable, we
quote Kibble:
\begin{quote}
``It is perfectly possible to describe our model without ever
introducing the notion of SSB, merely by writing down the
(phenomenological) Lagrangian. Indeed if the physical world were
described by this model, it is to the latter rather than to the former
to which we should be led by experiment. The only advantage of SSB is
that it is easier to understand the appearance of an exact symmetry
than an approximate one''~\cite{HonestGuy}.
\end{quote}
Such honesty is nowadays refreshing. It is all perhaps a matter of
taste. Tastes change over time, though; and to some the works of the
``exact'' symmetry are uglier than the refusal to deal with
unobservable fields.

\begin{acknowledgement}
MD was supported by the Deutsche Forschungsgemeinschaft through the
Institutional Strategy of the University of G\"ot\-tingen. JMG-B is
grateful to Luis J. Boya for calling attention to the
paper~\cite{NoClothesKing} and sound counsel on representations of the
classical Lie algebras. He is also indebted to Jean Zinn-Justin and
Marc Henneaux for illuminating conversations on the BRST invariance of
SSB models. His work was supported by DGIID--DGA (grant~E24/2). FS
thanks Stefan Tapprogge for valuable information about the
experimental Higgs sear\-ches. JCV acknowledges support from the
Vicerrector\'ia de Investigaci\'on of the University of Costa Rica.
\end{acknowledgement}


\begin{thebibliography}{79}
 
\bibitem{EB}
F. Englert and R. Brout,
Phys. Rev. Lett. \textbf{13} (1964)~321.%--323
%%CITATION = PRLTA,13,321;%%

\bibitem{H1}
P. W. Higgs,
Phys. Lett. \textbf{12} (1964)~132.%--133
%%CITATION = PHLTA,12,132;%%

\bibitem{H2}
P. W. Higgs,
Phys. Rev. Lett. \textbf{13} (1964)~508.%--509
%%CITATION = PRLTA,13,508;%%

\bibitem{GHK}
G. S. Guralnik, C. R. Hagen and T. W. B. Kibble,
Phys. Rev. Lett. \textbf{13} (1964)~585.%--587
%%CITATION = PRLTA,13,585;%%

\bibitem{PastRemembered}
G. S. Guralnik,
Int. J. Mod. Phys. A \textbf{24} (2009) 2601.%--2627
%%CITATION = IMPAE,A24,2601;%%
 
\bibitem{AH04}
I. J. R. Aitchison and A. J. G. Hey,
\textit{Gauge Theories in Particle Physics: QCD and the Electroweak 
Theory}
(IOP Publishing, Bristol, 2004).

\bibitem{Rafferty}
L. O'Raifeartaigh,
\textit{Group Structure of Gauge Theories}
(Cambridge University Press, Cambridge, 1986).

\bibitem{ModernosPragmaticos}
C. Burgess and G. Moore,
\textit{The Standard Model: a Primer}
(Cambridge University Press, Cambridge, 2007).

\bibitem{Okun}
L. B. Okun,
in \textit{Surveys in High Energy Physics} \textbf{5} (1986), p.~214.

\bibitem{TiniDixit}
M. J. G. Veltman,
Rev. Mod. Phys. \textbf{72} (2000) 341.%--349
%%CITATION = RMPHA,72,341;%%

\bibitem{TruthInRemembrance}
M. J. G. Veltman,
in \textit{The Rise of the Standard Model},
ed.~by L. Hoddeson, L. Brown, M. Riordan and M. Dresden
(Cambridge University Press, Cambridge, 1997), p.~145.%--178.

\bibitem{GFitter}
H. Fl\"acher, M. Goebel, J. Haller, A. Hoecker, K.~M\"onig and
J. Stelzer,
% ``Revisiting the global fit of the Standard Model and beyond with
% Gfitter'',
Eur. Phys.~J. C \textbf{60} (2009) 543.%--583.
%%CITATION = EPHJA,C60,543;%%

\bibitem{Cha-cha-cha}
M. S. Chanowitz,
Phys. Rev. D \textbf{66} (2002) 073002.%1--15
%%CITATION = PHRVA,D66,073002;%%

\bibitem{ELangacker08}
J. Erler and P. Langacker,
Acta Phys. Polon. B \textbf{39} (2008) 2595.%--2610
%%CITATION = APPOA,B39,2595;%%

\bibitem{GJWells08}
S. Gopalakrishna, S. Jung and J. D. Wells,
Phys. Rev. D \textbf{78} (2008) 055002.%%1--6
%%CITATION = PHRVA,D78,055002;%%

\bibitem{DarkSide}
H. Davoudiasl, R. Kitano, T. Li and H. Murayama,
Phys. Lett. B \textbf{609} (2005) 117.%--123
%%CITATION = PHLTA,B609,117;%%

\bibitem{Joraetal}
R. Jora, S. Moussa, S. Nasri, J. Schechter and M. Naeem Shahid,
Int. J. Mod. Phys. A \textbf{23} (2008) 5159.%--5172
%%CITATION = IMPAE,A23,5159;%%

\bibitem{Lightbeing}
F. Wilczek,
in \textit{Perspectives on LHC physics},
ed.~by G. Kane and A. Pierce
(World Scientific, Singapore,~2008).%p.~233--257.

\bibitem{AAh}
T. Aaltonen \textit{et al},
% ``Study of multi-muon events produced in $p\bar p$ collisions at
% $\sqrt s = 1.96$~TeV'',
hep-ex/0810.5357.
%%CITATION = ARXIV 0810.5357;%%

\bibitem{Ptochos}
F. Ptochos,
% ``Multi-muon events at the CDF'',
hep-ex/0907.0146.
%%CITATION = ARXIV 0907.0146;%%

\bibitem{Fermilab}
P. Giromini \textit{et al},
% ``Phenomenological interpretation of the multi-muon events reported
% by the CDF collaboration'',
hep-ph/0810.5730.
%%CITATION = ARXIV 0810.5730;%%

\bibitem{Straggler}
M. J. Strassler,
% ``Flesh and blood, or merely ghosts? Some comments on the multi-muon
% study at CDF'',
hep-ph/0811.1560.
%%CITATION = ARXIV 0811.1560;%%

\bibitem{Hamburg}
M. J. Strassler,
% ``New signatures and challenges for the LHC'',
% in Proceedings of the 38th International Symposium on Multiparticle
% Dynamics, p.~307; %--311
hep-ph/0902.0377.
%%CITATION = ARXIV 0902.0377;%%

\bibitem{NowD0Counterattacks}
V. M. Abazov \textit{et al},
% ``Evidence for an anomalous like-sign dimuon charge asymmetry'',
hep-ex/1005.2757.

\bibitem{Dobrescoso}
B. A. Dobrescu, P. J. Fox and A. Martin,
% ``CP violation in $B_s$ mixing from heavy Higgs exchange'',
hep-ph\slash 1005.4238.

\bibitem{DKS}
M. D\"utsch, F. Krahe and G. Scharf, 
%``Scalar QED Revisited'',
Nuovo Cim. A \textbf{106} (1993) 277.%--307
%%CITATION = NUCIA,A106,277;%%

\bibitem{ScharfQED}
G. Scharf,
\textit{Finite Quantum Electrodynamics. The Causal Approach}
(Springer, Berlin, 1995).

\bibitem{LuisDixit}
L. Alvarez-Gaum\'e and L. Baulieu,
Nucl. Phys. B \textbf{212} (1983) 255.%--267
%%CITATION = NUPHA,B212,255;%%

\bibitem{Pokorski}
S. Pokorski,
\textit{Gauge Field Theories}
(Cambridge University Press, Cambridge, 2000).

\bibitem{CLT74}
J. M. Cornwall, D. N. Levin and G. Tiktopoulos,
Phys. Rev. D \textbf{10} (1974) 1145.%--1167
%%CITATION = PHRVA,D10,1145;%%

\bibitem{LT75}
D. N. Levin and G. Tiktopoulos,
Phys. Rev. D \textbf{12} (1975) 415.%--420
%%CITATION = PHRVA,D12,415;%%

\bibitem{CabezondelaSal}
M. D\"utsch and B. Schroer,
J. Phys. A \textbf{33} (2000) 4317.
%%CITATION = JPAGB,33,4317;%%

\bibitem{Altabonazo}
H. Ruegg and M. Ruiz-Altaba,
Int. J. Mod. Phys. A \textbf{19} (2004) 3265.
%%CITATION = IMPAE,A19,3265;%%

\bibitem{Felicitas}
J. M. Gracia-Bond\'{\i}a,
% ``BRS invariance for massive boson fields'',
in \textit{Geometrical and Topological Methods for Quantum Field
Theory}, ed.~by H.~Ocampo, E.~Pari\-guan and S.~Paycha
(C.~U.~P., Cambridge, 2010), p.~220.%--252.
%%CITATION = ARXIV 0808.2853;%%

\bibitem{PCostello}
P. Costello,
% ``The mathematical structure of the quantum BRST constraint method'',
math.OA/0905.3570.
%%CITATION = ARXIV 0905.3570;%%

\bibitem{Scharf}
G. Scharf,
\textit{Quantum Gauge Theories: A True Ghost Story}
(Wiley, New York, 2001).
% See also the 2nd~edition (2010) at http://books.google.com/

\bibitem{DHKS}
M. D\"utsch, T. Hurth, F. Krahe and G. Scharf,
%Yang-Mills I-II
Nuovo Cim. A \textbf{106} (1993) 1029;%--1041
~\textit{ibidem} \textbf{107} (1994) 375.%--406 
%%CITATION = NUCIA,A106,1029;%%
%%CITATION = NUCIA,A107,375;%%

\bibitem{DHS}
M. D\"utsch, T. Hurth and G. Scharf, 
%Yang-Mills III-IV 
Nuovo Cim. A \textbf{108} (1995) 679;%--708
~\textit{ibidem} \textbf{108} (1995) 737.%--774
%%CITATION = NUCIA,A108,679;%%
%%CITATION = NUCIA,A108,737;%%

\bibitem{Tobyalone}
T. Hurth,
Ann. Phys. (NY) \textbf{244} (1995) 340.%--425
%%CITATION = APNYA,244,340;%%

\bibitem{PGI-EW-I}
M. D\"utsch and G. Scharf,
Ann. Phys. (Leipzig) \textbf{8} (1999) 359.%--387
%%CITATION = ANPYA,8,359;%%

\bibitem{PGI-EW-II}
A. Aste, G. Scharf and M. D\"utsch,
Ann. Phys. (Leipzig) \textbf{8} (1999) 389.%--404
%%CITATION = ANPYA,8,389;%%

\bibitem{PepinsFriend}
D. R. Grigore,
J. Phys. A \textbf{33} (2000) 8443.%--8746
%%CITATION = JPAGB,33,8443;%%

\bibitem{ESItalk}
R. Stora,
% ``Local gauge groups in quantum field theory: perturbative gauge
% theories'',
talk given at ESI, Vienna, 1997.

\bibitem{Michael05}
M. D\"utsch,
Ann. Phys. (Leipzig) \textbf{14} (2005) 438.%--461
%%CITATION = ANPYA,14,438;%%

\bibitem{FortunaJuvet}
M. Ambauen and G. Scharf,
% ``Violation of quantum gauge invariance in Georgi--Glashow $SU(5)$'',
hep-th/0409062.
%%CITATION = HEP-TH 0409062;%%

\bibitem{EpsteinGlaser} 
H. Epstein and V. Glaser, 
%``The role of locality in perturbation theory'', 
Ann. Inst. H. Poincar\'e \textbf{A 19} (1973) 211.%--295
%%CITATION = AHPAA,19,211;%%

\bibitem{QNC}
T. Hurth and K. Skenderis,
%``Quantum Noether method'',
Nucl. Phys. B \textbf{541} (1999) 566.%--614
%%CITATION = NUPHA,B541,566;%%

\bibitem{MWI}
M. D\"utsch and F.-M. Boas,
%``The Master Ward Identity''
Rev. Math. Phys. \textbf{14} (2002) 977.%--1049
%%CITATION = RMPHE,14,977;%%

\bibitem{MWIbis}
M. D\"utsch and K. Fredenhagen, 
%``The Master Ward Identity and generalized Schwinger--Dyson equation
%in classical field theory''
Commun. Math. Phys. \textbf{243} (2003) 275.%--314
%%CITATION = CMPHA,243,275;%%

\bibitem{QAP}
F. Brennecke and M. D\"utsch,
%``Removal of violations of the Master Ward Identity 
%in perturbative QFT'',
Rev. Math. Phys. \textbf{20} (2008) 119.%--172
%%CITATION = RMPHE,20,119;%%

\bibitem{BurningBurnel}
A. Burnel,
\textit{Noncovariant Gauges in Canonical Formalism}
(Springer, Berlin, 2009).

\bibitem{DF-QED}
M. D\"utsch and K. Fredenhagen,
%QED
Commun. Math. Phys. \textbf{203} (1999) 71.%--105
%%CITATION = CMPHA,203,71;%%

\bibitem{KO}
T. Kugo and I. Ojima,
Progr. Theor. Phys. Suppl. \textbf{66} (1979)~1.
%%CITATION = PTPSA,66,1;%%

\bibitem{BHH-I}
G. Barnich, M. Henneaux, T. Hurth and K. Skenderis,
%``Cohomological analysis of gauged-fixed gauge theories'',
Phys. Lett.~B \textbf{492} (2000) 376.%--384
%%CITATION = PHLTA,B492,376;%%

\bibitem{BHH-II}
G. Barnich, T. Hurth, K. Skenderis,
%``Comments on the gauge fixed BRST cohomology and the quantum Noether
%method'',
Phys. Lett.~B \textbf{588} (2004) 111.%--118
%%CITATION = PHLTA,B588,111;%%

\bibitem{SecundumMathew}
P. M. Mathews, M. Seetharaman and M. T. Simon,
% ``Indecomposability of Poincar\'e group representations over
% massless fields, and the quantization problem for electromagnetic
% potentials'',
Phys. Rev. D \textbf{9} (1974) 1706.%--1710
%%CITATION = PHRVA,D9,1706;%%

\bibitem{BS1}
B. Schroer,
% ``Jorge A. Swieca's contributions to quantum field theory in the 60s
% and 70s and their relevance in present research'',
Eur. Phys. J. H \textbf{35} (2010) 53.%--88
% physics.hist-ph/0712.0371.
%%CITATION = ARXIV 0712.0371;%%

\bibitem{BS2}
B. Schroer,
% ``Unexplored regions in QFT: delocalization of quantum matter through 
% interactions with zero mass potentials'',
hep-th/1006.3543.
%%CITATION = ARXIV 1006.3543;%%

\bibitem{Oldhand}
A. Burnel,
Acta Phys. Polon. B \textbf{27} (1996) 2441.%--2451
%%CITATION = APPOA,B27,2441;%%

\bibitem{Almasgemelas}
M. Chaichian and K. Nishijima,
Eur. Phys.~J. C \textbf{22} (2001) 463.%--477.
%%CITATION = EPHJA,C22,463;%%

\bibitem{AncaN}
K. Nishijima and A. Tureanu,
Eur. Phys.~J. C \textbf{53} (2008) 649.%--657.
%%CITATION = EPHJA,C53,649;%%

\bibitem{DogDine}
M. Dine,
\textit{Supersymmetry and String Theory. Beyond the Standard Model}
(C.~U.~P., Cambridge, 2007).

\bibitem{SmokeScreen}
M. J. G. Veltman,
%``The screening theorem and the Higgs system''
Acta Phys. Polon. B \textbf{25} (1994) 1627.%--1635
%%CITATION = APPOA,B25,1627;%%

\bibitem{Ausonia}
J. M. Gracia-Bond\'{\i}a,
%``On the causal gauge principle'',
%to appear in the Proceedings of the workshop at MPIM~Bonn
hep-th/0809.0160.
%%CITATION = ARXIV 0809.0160;%%

\bibitem{Tini75}
M. J. G. Veltman,
%``Cosmology and the Higgs mass'',
Phys. Rev. Lett. \textbf{34} (1975) 777.%--777
%%CITATION = PRLTA,34,777;%%

\bibitem{Srednicki}
M. Srednicki,
\textit{Quantum Field Theory}
(Cambridge University Press, Cambridge, 2007).

\bibitem{GGstrikeagain}
H. Georgi and S. L. Glashow,
%``Unity of all elementary particle forces'',
Phys. Rev. Lett. \textbf{32} (1974) 438.%--441
%%CITATION = PRLTA,32,438;%%

\bibitem{pseudoWeinberg}
S. Weinberg,
Phys. Rev. Lett. \textbf{29} (1972) 1698.%--1701
%%CITATION = PRLTA,29,1698;%%

\bibitem{VacuumCleaner}
S. Weinberg,
\textit{The Quantum Theory of Fields II}
(Cambridge University Press, Cambridge, 1996).

\bibitem{CoLo}
A. Connes and J. Lott,
Nucl. Phys. B (Proc. Suppl.) \textbf{18} (1990) 29.%--47
%%CITATION = NUPHZ,18,29;%%

\bibitem{CoEFSch}
R. Coquereaux, G. Esposito-Far\`ese and F. Scheck,
Int. J. Mod. Phys. A \textbf{7} (1992) 6555.%--6593
%%CITATION = IMPAE,A7,6555;%%

\bibitem{Zappafrank}
H. Figueroa, J. M. Gracia-Bond\'{\i}a, F. Lizzi and J. C. V\'arilly,
%``A nonperturbative form of the spectral action principle in
%noncommutative geometry''
J. Geom. Phys. \textbf{26} (1998) 329.%--339
%%CITATION = JGPHE,36,329;%%

\bibitem{ConnesMix}
A. H. Chamseddine, A. Connes and M. Marcolli,
Adv. Theor. Math. Phys. \textbf{11} (2007) 991.%--1089
%%CITATION = 00203,11,991;%%

\bibitem{Heil-Eins}
A. Heil, N. A. Papadopoulos, B. Reifenh\"auser and F. Scheck,
Nucl. Phys. B \textbf{293} (1987) 445.%--460
%%CITATION = NUPHA,B293,445;%%

\bibitem{Heil-Zwei}
A. Heil, A. Kersch, N. A. Papadopoulos, B. Reifenh\"auser and F.
Scheck,
Ann. Phys. (NY) \textbf{200} (1990) 206.%--215
%%CITATION = APNYA,200,206;%%

\bibitem{ChacraRanaLoco}
J. L. Chkareuli, C. D. Froggatt and H. B. Nielsen,
Phys. Rev. Lett. \textbf{87} (2001) 091601.%1--4
%%CITATION = PRLTA,87,091601;%%

\bibitem{Oldebook}
F. Scheck,
\textit{Leptons, Hadrons and Nuclei}
(North-Holland, Amsterdam, 1983).

\bibitem{NoClothesKing}
H. Lyre,
Intl. Studies Philos. Sci. \textbf{22} (2008) 119.%--133
%%CITATION = 00249,22,119;%%

\bibitem{HombreOreja}
J. Earman,
Intl. Studies Philos. Sci. \textbf{18} (2004) 173.%--198
%%CITATION = 00249,18,173;%%

\bibitem{HonestGuy}
T. W. B. Kibble,
Phys. Rev. \textbf{155} (1967) 1554.%--1561
%%CITATION = PHRVA,155,1554;%%


\end{thebibliography}
\end{document}